\documentclass[11pt]{article}

\usepackage{calc}
\makeatletter
\@addtoreset{equation}{section}

\makeatother
\usepackage{feynmf}
\usepackage{graphicx}
\usepackage{amssymb,amsfonts,amsmath}

\def\beq{\begin{equation}}
\def\eeq{\end{equation}}
\def\bea{\begin{eqnarray}}
\def\eea{\end{eqnarray}}

\newcommand{\vph}{\varphi}

\newcommand{\non}{\nonumber \\}

\newcommand{\hp}{\hat{p}}

\begin{document}
%%%%%%%%%%%%%%%%%%%%%%%%%%%%%%%%%%%%%%%%%%%%%%%%%%%%%%%%%%%%%%

%% -----------------------------------------------------------
%% title, etc.
\renewcommand{\thefootnote}{\fnsymbol{footnote}}
\newcommand{\inst}[1]{\mbox{$^{\text{\textnormal{#1}}}$}}
%% ----- preprint numbers -----
\begin{flushright}
DFTT 05/2010\\
EPHOU 10-002\\
June, 2010
\end{flushright}
\mbox{}\\\bigskip\bigskip
\begin{center}
%% ----- title -----
{\LARGE Species Doublers as Super Multiplets
in Lattice Supersymmetry: \\
Exact Supersymmetry with Interactions for $D=1$ $N=2$}\\[8ex]
%
%% ----- author -----
{\large
Alessandro D'Adda\inst{a}\footnote{\texttt{dadda@to.infn.it}},
Alessandra Feo\inst{a}\footnote{\texttt{feo@to.infn.it}},
Issaku Kanamori\inst{a}\footnote{\texttt{kanamori@to.infn.it}},
Noboru Kawamoto\inst{b}\footnote{\texttt{kawamoto@particle.sci.hokudai.ac.jp}}\\[1ex]
{\normalsize and}
Jun Saito\inst{b}\footnote{ \texttt{saito@particle.sci.hokudai.ac.jp}}. }\\[4ex]
%
%% ----- institute -----
{\large\itshape
\inst{a} INFN Sezione di Torino, and\\
Dipartimento di Fisica Teorica,
Universita di Torino\\
I-10125 Torino, Italy\\[3ex]
\inst{b} Department of Physics, Hokkaido University\\
Sapporo, 060-0810 Japan}
\end{center}
\bigskip\bigskip
\setcounter{footnote}{0}
\renewcommand{\thefootnote}{\arabic{footnote}}
%% -----------------------------------------------------------

%% -----------------------------------------------------------
%% abstract
\begin{abstract}
We propose a new lattice superfield formalism in momentum
representation which accommodates species doublers of the lattice
fermions and their bosonic counterparts as super multiplets. We
explicitly show that one dimensional $N=2$ model with interactions
has exact Lie algebraic supersymmetry on the lattice for all super
charges.

In coordinate representation the finite difference operator is
made to satisfy Leibnitz rule by introducing a non local  product,
the ``star'' product,  and the exact lattice supersymmetry is
realized. The standard momentum conservation is replaced on the
lattice by the conservation of the sine of the momentum, which plays a
crucial role in the formulation. Half lattice spacing structure is
essential for the one dimensional model and the lattice
supersymmetry transformation can be identified as a half lattice
spacing translation combined with alternating sign structure.
Invariance under finite translations and locality in the continuum
limit are explicitly investigated and shown to be recovered.
Supersymmetric Ward identities are shown to be satisfied at one
loop level. Lie algebraic lattice supersymmetry algebra of this
model suggests a close connection with Hopf algebraic exactness of
the link approach formulation of lattice supersymmetry.

\end{abstract}
%% -----------------------------------------------------------

% lattice gauge theory,
% supersymmetry
% field theories in dimensions other than four,


%% Los Alamos Database Number:

\section{Introduction}

If we regularize massless fermions naively on a lattice, it is
unavoidable that species doublers appear. Since massless particles
cannot be put in the  rest frame by means of a Lorentz
transformation, helicity or chirality cannot be changed with a
momentum change, while  species doublers in different momentum
region may have different helicity. Therefore species doublers
have to be considered as different particles~\cite{No-Go}.
However, species doublers of chiral fermions on a lattice are
usually considered as unwanted particles, the so called doubling problem,
although the enlarged
degree of freedom (d.o.f.) is customarily identified as a flavor
(taste) d.o.f..

The equivalence of the above naive fermion formulation and the
staggered fermion formulation can be shown by a spin
diagonalization procedure~\cite{K-S} and the staggered fermion can
be transformed into the Kogut-Susskind type fermion
formulation~\cite{Kogut-Susskind} by considering double size
lattice structure, where the flavor d.o.f. was identified~\cite{
K-M-N}. This double size structure makes it possible to have a
correspondence with differential forms and then the equivalence of
the staggered fermion and Dirac-K\"ahler fermion on the lattice
can be proved by introducing a noncommutativity between
differential forms and fields~\cite{K-K}. Therefore all these
lattice fermion formulations are exactly equivalent.

%The above naive fermion formulation can be shown to be equivalent to
%Dirac-K\"ahler fermion formulation via the equivalence with the staggered
%fermion formulation.

In the link approach of lattice supersymmetry~\cite{DKKN}, the
super charges are expanded on the basis of Dirac matrices by the
Dirac-K\"ahler twisting procedure~\cite{KKU}. The corresponding
d.o.f. of the super charges are then exactly the same as those of
fermionic species doublers and the geometric correspondence
between the particles as species doublers and super multiplets is
expected from the equivalence of the naive fermion and
Dirac-K\"ahler fermion formulations on the lattice. In other words
the species doublers are necessary fields to construct the super
multiplets of extended supersymmetry: $N=2$ in two dimensions and
N=4 in four dimensions. The flavor (taste) d.o.f. of chiral
fermions are thus expected to be identified as extended
supersymmetry d.o.f.. In this paper we explicitly show how the
species doublers for both fermions and bosons can be identified as
super multiplets of extended supersymmetry for the simplest model
of $N=2$ in one dimension. We propose to introduce lattice counter
parts of bosonic and fermionic ``superfields'' where species
doublers are accommodated.

The one dimensional $N=2$ model was proposed as a supersymmetric quantum
mechanics by Witten~\cite{Witten} and the lattice version has been
investigated by several authors~\cite{C-G,Giedt} as the simplest
model to clarify the fundamental problems of lattice
supersymmetry. It was shown that a species doubler of the Wilson
fermion term, having a mass proportional to the inverse lattice
constant, breaks supersymmetry and  a bosonic counter term is
needed to remove the unwanted contribution~\cite{Giedt}. Numerical
evaluation of boson and fermion masses show that they approach the
same value in the continuum limit, suggesting the recovery of
supersymmetry, only when the counter term is
introduced~\cite{C-G,CKU}. In this model the species doubler
interferes with supersymmetry and its influence has to be removed
by the bosonic counter term. It has also been recognized that only
one exact supersymmetry of the type $Q^2=0$, which can be
identified as the scalar part of a twisted supersymmetry for this
supersymmetric quantum mechanics, is realized when  interaction
terms are included~\cite{CKU}.
This system also provides a nice arena for a numerical method
for detecting spontaneous supersymmetry breaking \cite{KSS}.

One dimensional $N=2$ lattice supersymmetric model of this paper is
constructed in parallel to the continuum superspace
formulation~\cite{Cooper:1982dm,Dutch,Arianos:2008ai} and is slightly different
from the supersymmetric quantum mechanics model.
%However the basic
%structure of the lattice nature is the same with the supersymmetric
%quantum mechanics.
The lattice model we propose in this paper is exactly supersymmetric
for two supersymmetry charges even with interaction terms and the
counter term is not necessary to fulfill Ward-Takahshi identity
since the bosonic and fermionic species doublers are identified as
physical particles in supermultiplets.  In the momentum
representation this model has, however, a lattice counter part of
trigonometric momentum conservation, which was first proposed by
Dondi and Nicolai~\cite{D-N} in the very first paper of lattice
supersymmetry. In the coordinate space we introduce a new star
product which makes the lattice difference operator satisfy
Leibniz rule and then the exact lattice supersymmetry is realized.
The model has mildly nonlocal interactions which approach local
interactions in the continuum limit.

There is a long history of attempts to realize exact sypersymmetry
on a lattice. See \cite{CKU,Feo:2002yi} for some  references. However
exact lattice supersymmetry with interactions for full extended
supersymmetry has never been realized except for the nilpotent
super charge. This difficulty is essentially related to the lattice
chiral fermion problem and to the breakdown of the Leibniz rule for
the lattice difference operator. Instead of formulating an exact
supersymmetry for local interactions in the coordinate space,
there has been several attempts that approach the problem from
momentum representation point of view~\cite{Hanada:2007ti,
Kadoh:2009sp, Bergner:2009vg}. This may be related to the
following  claim: if one tries to include the difference operator
in a supersymmetry algebra, one cannot avoid  introducing nonlocal
interactions~\cite{Kato:2008sp}. The momentum representation can
take care the nonlocal nature of the formulation. In this paper we
first establish a formulation  of exact lattice supersymmetry in
the momentum representation. Then we reformulate the momentum
space version into the coordinate space by introducing a new non
local ``star'' product.

In the link approach for lattice
supersymmetry~\cite{DKKN} the claim that
exact lattice supersymmetry had been realized for all super
charges was questioned by several authors~\cite{Dutch,B-C-K}. In
fact it was stressed that all the extended supersymmetry are
broken when the shift parameter of the scalar super charge is non
zero $a\ne 0$~\cite{D-M} while it exactly coincides with the
orbifold construction of lattice supersymmetry when
$a=0$~\cite{Kaplan:2003,catterall,sugino}.  It was however shown later
on that lattice supersymmetry can be formulated consistently
within the framework of a Hopf algebra, accounting for the
breaking  of the Leibniz rule for the difference operator, and of
the  mild noncommutativity between fields carrying  a shift. Thus
exact lattice supersymmetry holds within the Hopf algebraic
symmetry~\cite{DKS}.

One of the important aims of the present paper is to clarify the
fundamental nature of the link approach within a simple
one-dimensional model. Since we find  a new exact lattice
supersymmetry formulation, it would be interesting to compare the
algebraic structure of the link approach with this new
formulation.
We point out the interesting possibility, supported by several arguments, that
the star product formulation of current model and the link
approach formulation are equivalent.

This paper is organized as follows: We explain the basic ideas of
the formulation in section 2. We then briefly explain the
continuum version of the model which we investigate in this paper
in section 3. Then it will be explained in section 4 how the
species doublers can be naturally accommodated into supersymmetry
transformations together with trigonometric momentum conservation.
In section 5 an exact supersymmetry invariant action with
interaction terms in momentum representation on the infinite
lattice will be proposed. In section 6 the recovery of the
translational invariance of this model in the continuum limit is
investigated. It will be confirmed that supersymmetric
Ward-Takahashi identities are satisfied. In section 7 we propose a
new star product which makes lattice difference operator satisfy
Leibniz rule and exact lattice supersymmetry be realized in the
coordinate space. A close connection with the link approach will
be discussed. We then summarize the result of this paper and
discuss remaining problems in the final section. In the appendix
1-loop radiative corrections of propagators in Ward-Takahashi
identity are summarized.

\section{Basic ideas}
\label{basic-ideas}

\begin{figure}
 \hfil
 \includegraphics{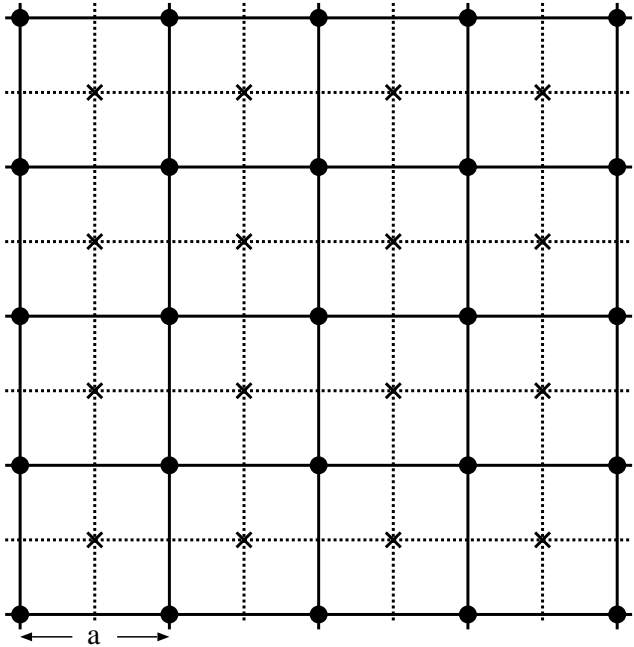}
  \caption{$\Phi_1(\vec{x})$ takes the same value $c_1$ in both $\bullet$ and $\times$ points, while $\Phi_2(\vec{x})$ takes the value $+c_2$ in the $\bullet$ points and $-c_2$ in the $\times$ points. Adjoining $\bullet$ and $\times$ points
  are joined by the fermionic links which are not shown in the figure.}
 \label{fig:lattice}
\end{figure}

One of the most distinctive features of the so called link
approach to lattice supersymmetry is the introduction in place of
the usual hyper cubic lattice of extended lattices that account for
the underlying supersymmetry algebra. The idea is the following:
on the lattice infinitesimal translations are replaced by finite
displacements, or shifts, represented typically for an hyper cubic
lattice by orthogonal vectors $\vec{n}_\mu$ of length equal to the
lattice spacing $a$. The vectors $\vec{n}_\mu$ generate the whole
lattice and each point of the lattice can be reached from a given
point by means of a finite number of such displacements. It
follows that a translationally invariant field configuration, or
vacuum, is a constant field configuration on the lattice. In the
link approach a shift $\vec{a}_A$ is  associated also to each
supersymmetry charge $Q_A$, in the same way as $\vec{n}_\mu$ is
associated to the generator $P_\mu$ of translations . The shifts
$\vec{a}_A$ are not arbitrary, but they are constrained by the
supersymmetry algebra: consistency with the algebra requires that
the constraint $\vec{a}_A + \vec{a}_B = \pm \vec{n}_\mu$ must be
satisfied for each non-vanishing anticommutator $\{ Q_A,Q_B\} =
P_\mu$ of the superalgebra. Only a limited number of supersymmetry
algebras, in particular the $N=2$  SUSY algebra in $2$ dimensions
and the $N=4$ SUSY algebra in $4$ dimensions, are compatible with
these constraints. The extended lattices introduced in the link
approach are generated by the displacements $\vec{a}_A$  and
$\vec{n}_\mu$ and hence contain, with respect to the standard
hyper cubic lattices, new  links of the type $ (\vec{x} , \vec{x} +
\vec{a}_A )$ which we will call ``fermionic'' and new points. The
key remark now is that not all points of the extended lattices can
be reached from a given one simply by translations. An extended
lattice will consist in general of more copies of the hyper cubic
lattice connected by ``fermionic'' links, and translations will only
make us move within each copy. So for a field configuration to be
invariant under translations it is not necessary to be constant
over the whole extended lattice but only separately over each copy
of the hyper cubic lattice. In other words the number of field
configurations which are invariant under translations in an
extended lattice is equal to the number of hyper cubic sublattices
contained in it. Consider as an example the $N=2$ superalgebra in
$2$ dimensions. This superalgebra contains four supersymmetry
charges $Q_A$ besides the generators of translations in two
dimensions. In its discrete lattice version, described in detail
in ref ~\cite{D'Adda:2004jb}, four shifts $\vec{a}_A$ are
associated to the four supesymmetry charges, constrained by the
non-vanishing anticommutators of the superalgebra, as discussed
above. The constraints do not determine $\vec{a}_A$ completely:
one of them is arbitrary. However if we further require the
resulting extended lattice to be  invariant under $\frac{\pi}{2}$
rotations, then the solution is unique and given by \beq \vec{a}_A
= (\pm \frac{a}{2}, \pm \frac{a}{2}), \label{Aa} \eeq where $a$
denotes the lattice spacing. The vectors (\ref{Aa}), together with
the shifts associated to translations, namely $\vec{n}_1 = (a,0)$
and $\vec{n}_2 = (0,a)$, generate the extended lattice of the
$N=2$, $D=2$ SUSY algebra. This is shown in fig. (\ref{fig:lattice}).
The points $\vec{x}$ of the lattice
have coordinates: \beq \vec{x} = ( \frac{n a}{2}, \frac{m a}{2} ),
\label{vecx} \eeq where  $n + m$ is even ( $n$ and $m$ both even
or both odd).

The extended lattice is then made by two copies of
the cubic lattice, in fact by the original lattice ( $n$ and $m$
even) and its dual ($n$ and $m$ odd) connected by the ``fermionic''
links $(\vec{x}, \vec{x}+\vec{a}_A)$. It is clear  that there are
two independent field configurations on the extended lattice that
are invariant under translation, namely: \beq \Phi_1(\vec{x}) =
c_1~~~~~~~~~~~,~~~~~~~~~~\Phi_2(\vec{x}) = (-1)^{n} c_2,
\label{vcconf1} \eeq with $c_1$ and $c_2$ constant.

It is
interesting to note that this phase remind us of the phase of
staggered fermion~\cite{K-S} This is relevant because we expect
the degrees of freedom of the theory in the continuum limit to be
associated to small fluctuations around translationally invariant
vacua, so that  fluctuations around the two configurations of eq.
(\ref{vcconf1}) will correspond to two distinct degrees of freedom
in the continuum limit. Hence one degree of freedom on the
extended lattice will absorb two degrees of freedom of the
continuum theory. This fact, namely that the extended lattice
implies a correspondingly reduced number of independent fields,
was not fully appreciated in the original formulation of the link
approach and is one of the key points of the present paper.
 It is clear that since the different copies of the hyper cubic lattice in an extended lattice are generated by the extra links associated to supersymmetry charges it is natural to expect  the different continuum degrees of freedom associated to a single lattice degree of freedom  to be part of a supersymmetric multiplet.
 It is instructive to look at the previous example from the point of view of the momentum space representation, which will play a crucial role in what follows. The first Brillouin zone associated to the lattice  (\ref{vecx}) is the square  defined by  $-\frac{2 \pi}{a} \leq p_1+p_2 \leq \frac{2 \pi}{a}$ and $-\frac{2 \pi}{a} \leq p_1-p_2 \leq \frac{2 \pi}{a}$. That means that in the momentum space fields will be periodic with period  $\frac{4 \pi}{a}$ in the variables $p_1 +p_2$ and $p_1 - p_2$. The translationally invariant field configurations (\ref{vcconf}) correspond respectively to the center of the square, namely $\vec{p}=0$, and to the four vertices (which are all equivalent due to the periodicity), namely $ \vec{p}= (\pm \frac{2 \pi}{a},0)$ or  $ \vec{p}= (0,\pm \frac{2 \pi}{a})$. The latter is exaclty the species doubler.
 A solution of the fermion doubling problem becomes now possible. Fermion doubling originates from the fact that the fermionic kinetic term has a simple zero at $\vec{p}=0$ and hence has to vanish somewhere else in the Brillouin zone due to the lattice periodicity. Within the extended lattice scheme if the second zero occurs in correspondence of the other translationally invariant vacuum the would be doubler can be interpreted as a physical field, in fact as a supersymmetric partner of the original one at $\vec{p}=0$.

 We have used the example of the extended lattice of the $N=2$, $D=2$ supersymmetry to illustrate the ideas that we are going to develop in this paper. However the explicit example that we are going to study is a simpler one, $N=2$ supersymmetric model in one dimension ($D=1$).
 Before going into that we are going to consider an even simpler example, that has been considered in the present context in ref. ~\cite{D'Adda:2009es}.
 This is a one dimensional model with an $N=1$ supersymmetry.
 It is described in terms of a superfield:
\beq \Phi(x,\theta)= \varphi(x) + i \theta \psi(x), \label{superfield-0}\eeq
with a supersymmetry charge given by:
 \beq Q = \frac{\partial}{\partial \theta} +
i \theta \frac{\partial}{\partial x}, \label{susycharge}\eeq and
\beq Q^2 = i \frac{\partial}{\partial x}. \label{qsquare}\eeq On
the lattice derivatives are replaced by finite shifts of length
$a$, hence consistency with the algebra (\ref{qsquare}) requires
that the supercharge $Q$ is associated to a  shift $\frac{a}{2}$
. The extended lattice is then a one dimensional lattice with
spacing $\frac{a}{2}$, which can be thought of as the
superposition of two lattices with spacing $a$ each invariant
under translations and separated by a an $\frac{a}{2}$ shift
associated to the SUSY charge. Again there are two field
configurations invariant under translations, namely: \beq
\Phi_1(x) = c_1~~~~~~~~~~~,~~~~~~~~~~\Phi_2(x) = (-1)^{\frac{2
x}{a}} c_2 \label{vcconf} \eeq with $x=\frac{n a}{2}$. According
to the previous discussion fluctuations around $\Phi_1(x)$ and
$\Phi_2(x)$ will describe in the continuum limit two distinct
degrees of freedom. However we  have just two degrees of freedom
in this model, one bosonic and one fermionic, so the natural thing
is to associate the bosonic degree of freedom to fluctuations
around $\Phi_1(x)$ and the fermionic one to fluctuations around
$\Phi_2(x)$.

To understand the origin of the hyper lattice structure of half
lattice step and the alternating sign states, let us look at the
algebraic relation of the superfield and the supercharge from a
matrix point of view~\cite{Arianos:2008ai}. We now introduce the
following matrix form of the super coordinate and its derivative
as: \beq \theta=
\begin{pmatrix}
0 & 1  \\
0 & 0
\end{pmatrix}, ~~~~~~~~~~~~~~~~~
\frac{\partial}{\partial\theta}=
\begin{pmatrix}
0 & 0  \\
1 & 0
\end{pmatrix},
\eeq
which satisfy the following anticommutation relation:
\beq
\{\frac{\partial}{\partial\theta}~,~\theta\}~=~1.
\eeq
We may consider this matrix structure as an internal structure of the space time
coordinate. With respect to this internal structure the boson $\varphi$ is considered
as a field which commutes with $\theta$ and $\frac{\partial}{\partial\theta}$ and the
fermion $\psi$ as a field which anticommutes with  them. The component fields of
boson and fermion with respect to this internal structure has then the following form:
\beq
\varphi (x)=
\begin{pmatrix}
\varphi (x) & 0  \\
0 & \varphi (x)
\end{pmatrix}, \label{boson-matrix}
\eeq
\beq
\psi (x)=
\begin{pmatrix}
\psi (x) & 0  \\
0 & -\psi (x)
\end{pmatrix}.
\label{fermion-matrix} \eeq In the matrix formulation of fields
the coordinate dependence can be introduced by diagonal entries of
a big matrix as direct product to the internal matrix
structure~\cite{Arianos:2008ai}.  From the degrees of freedom
point of view for N=1 model in one dimension two fields of boson
and fermion on the same lattice site have this internal matrix
structure (\ref{boson-matrix}) and (\ref{fermion-matrix}). If we
consider the $N=2$ model in one dimension, which we will consider
later in this paper,  there are four independent fields on one
site we may then consider that the four fields of internal matrix
structure may be identified with the four independent d.o.f. of
fields.

We now ask a question: ``How do we interpret this internal space time structure
on the lattice ?'' A natural identification is an introduction of a half lattice step
structure to make a correspondence with two translational invariant states
in (\ref{vcconf}).  We can then identify a constant field of $\varphi(x)$ in
(\ref{boson-matrix}) as $\Phi_1(x)$ in (\ref{vcconf}) and a constant field
of $\psi(x)$ in (\ref{fermion-matrix}) as $\Phi_2(x)$ in  (\ref{vcconf}).

One can then write a lattice ``superfield'' corresponding to (\ref{superfield-0}) as
\beq
\Phi(x) = \varphi(x) + \frac{1}{2} (-1)^\frac{2x}{a} \psi(x) \label{lattsup},
\eeq
where we have introduced a factor $\frac {1}{2}$ for later convenience and taken away
the factor $i$ since the second term is not a product of two Grassmann numbers
to keep hermiticity. Since $\theta$ and $\frac{\partial}{\partial\theta}$ are
not Hermitian by them self in this matrix representation we have to take care the
hermiticity separately.
It is crucial to recognize at this stage that the  super coordinate structure
and fermionic nature of $\psi$ can be accommodated by the alternating sign
factor of half lattice spacing if this simple lattice representation works as a superfield.
We now introduce a matrix form of a fermionic super parameter by
\beq
\alpha=
\begin{pmatrix}
\alpha & 0  \\
0 & -\alpha
\end{pmatrix}, \label{alpha-matrix}
\eeq
where $\alpha$ is a Grassmann odd parameter. This parameter can then be expressed as
$\alpha (-1)^\frac{2x}{a}$ in accordance with the representation of the lattice
superfield in (\ref{lattsup}).

We now propose lattice supersymmetry transformations as a finite difference over a half
lattice spacing  $\frac{a}{2}$:
\beq
\delta \Phi(x) = a^{-\frac{1}{2}} \alpha (-1)^\frac{2x}{a}
\left( \Phi(x+\frac{a}{2}) - \Phi(x)\right). \label{suslatt}
\eeq
 In terms of the component fields the supersymmetry transformations (\ref{suslatt}) are:

 \bea &\delta \varphi(x) = -\frac{\alpha}{2}
\bigg[ \psi(x+\frac{a}{2}) + \psi(x) \bigg] \xrightarrow[a\to 0]{}%\to_{a\to 0}
-\alpha \psi(x) \, ,\label{suslatt1} \\
& \delta \psi(x) = 2 a^{-1} \alpha \bigg[ \varphi(x+\frac{a}{2})
- \varphi(x)  \bigg]  \xrightarrow[a\to 0]{}%\to_{a\to 0}
\alpha \frac{\partial
\varphi(x)}{\partial x} \, . \label{suslatt2} \eea
It is surprising that the half lattice translation together with alternating sign structure
(staggered phase) for the lattice superfields generates a correct lattice supersymmetry
transformation. We consider that this observation is a key of our formulation.

Although supersymmetry transformations (\ref{suslatt1}) and
(\ref{suslatt2}) have the correct structure, they violate hermiticity:
a factor $i$ is missing at the l.h.s. of (\ref{suslatt1}).
In order to restore the hermiticity of the supersymmetry
transformations symmetric finite differences must be used,
introducing a shift of $\frac{a}{4}$ of the fermionic fields sites
with respect to the bosonic ones. Hence, instead of writing the
superfield on the lattice as in (\ref{lattsup}) we shall introduce
$\Phi(x)$, with $x=n \frac{a}{4}$, defined by:
 \beq \Phi(x) = \left\{ \begin{array}{lc} & \varphi(x)~~~~~~~~\textrm{for}~~~
 x=n a/2 ,\\&\frac{1}{2} a^{1/2} e^{\frac{2i \pi x}{a}}
 \psi(x) ~~~\textrm{for}~~~x=(2n+1)a/4 .
 \end{array} \right. \label{spf} \eeq
 Again the supersymmetry transformations can be written in terms
 of $\Phi(x)$:
 \beq
 \delta\Phi(x) = \alpha a^{-1/2} e^{\frac{2 i \pi x}{a}} \left[
 \Phi(x+a/4) - \Phi(x-a/4) \right]. \label{st}
 \eeq
 By separating $\Phi(x)$ into its component fields according to
 (\ref{spf}) we find:
 \bea
&&\delta \varphi(x) = \frac{i \alpha}{2} \bigg[
\psi(x+\frac{a}{4}) + \psi(x-\frac{a}{4}) \bigg] \xrightarrow[a\to 0]{}%\to_{a\to 0}
i \alpha \psi(x) \, ,\label{suslattf1} \\
&& \delta \psi(x) = 2 a^{-1} \alpha \bigg[ \varphi(x+\frac{a}{4})
- \varphi(x-\frac{a}{4})  \bigg]  \xrightarrow[a\to 0]{}%\to_{a\to 0}
\alpha
\frac{\partial \varphi(x)}{\partial x} \, , \label{suslattf2} \eea
where $x$ is an even multiple of $a/4$ in (\ref{suslattf1}) and an
odd one in (\ref{suslattf2}). As in the continuum case the
commutator of two SUSY transformation is a translation, namely, on
the lattice, a finite difference of spacing $a$. For instance we
have for $\varphi(x)$ (the same applies to $\psi(x)$):
 \beq
 \delta_{\beta}\delta_{\alpha}\varphi(x) -
\delta_{\alpha}\delta_{\beta}\varphi(x) = \frac{2 i \alpha \beta}{a}
\left[ \varphi(x+a/2)-\varphi(x-a/2) \right]. \label{trs}
\eeq
It is instructive to look at the supersymmetry transformations given above from
the point of view of of the momentum space representation.
Let us consider first the Fourier transform of the component
fields $\psi(x) $ and $\varphi(x)$, and denote them by $\tilde
\psi(p)$ and $\tilde \varphi(p)$ respectively. The lattice spacing
being $a/2$, the Brillouin zone extends over a $\frac{4\pi}{a}$ interval with the two vacua
$\Phi_1(x)$ and $\Phi_2(x)$ corresponding respectively to $p=0$ and $p=\frac{2 \pi}{a}$.
The periodicity conditions are:
\beq \tilde
\varphi(p+\frac{4\pi}{a})=\tilde \varphi(p),\,\,\,\,\,\,\,\tilde
\psi(p+\frac{4\pi}{a})=-\tilde \psi(p), \label{period-0} \eeq
where
the minus sign in the case of $\tilde \psi$ is due to the $a/4$
shift in coordinate space of the fermionic field. The supersymmetry transformations
(\ref{suslattf1}) and (\ref{suslattf2}) are then given by:
\bea
&& \delta \tilde \varphi(p) = i \cos \frac{a p}{4} \, \alpha \tilde \psi(p)\label{susymom1}, \\
&& \delta \tilde \psi(p) = -i\frac{4}{a} \sin\frac{a p }{4}\,
\alpha \tilde \varphi(p).  \label{susymom2}
\eea
Eqs. (\ref{susymom1}) and (\ref{susymom2}) are consistent with both the periodicity conditions
(\ref{period-0}) and with the reality conditions expressed in momentum space by:
$ \tilde \varphi(p)^\dag = \tilde \varphi(-p)$ and
$\tilde\psi(p)^\dag =\tilde\psi(-p)$.
A more extensive analysis of the $D=1$, $N=1$ model, including the lattice action, can be found in
~\cite{D'Adda:2009es}. The point we want to emphasize here is the following: in order to derive the supersymmetry transformations (\ref{suslattf1}) and (\ref{suslattf2}), or equivalently their momentum space representation (\ref{susymom1}) and (\ref{susymom2}), we started from a bosonic field $\varphi(x)$ and a fermionic one $\psi(x)$ interpreted respectively as fluctuations around $\Phi_1(x)$ (i.e. $p=0$) and $\Phi_2(x)$ (i.e $p=\frac{2 \pi}{a}$). Either these two fields represent on the lattice a single degree of freedom whose statistic changes from bosonic to fermionic as the momentum moves from zero to $\frac{2 \pi}{a}$ ( and we don't know how to implement that consistently) or each field has a doubler with the same statistic in correspondence of the other vacuum. In this case the system will contain two bosonic and two fermionic degrees of freedom in the continuum, for which there is no room in the $D=1$, $N=1$ supersymmetry. Furthermore the action of this model is fermionic and thus the vacuum is not well defined.
We will show in the following section that it actually provides a consistent formulation of the $D=1$, $N=2$ supersymmetry , whose algebra contains a bosonic field $\varphi$ and a bosonic auxiliary  field $D$, described by the lattice bosonic field at $p=0$ and $p=\frac{2 \pi}{a}$ respectively and, in the fermionic sector, two  fields $\psi_1$ and $\psi_2$ described on the lattice by the fluctuations of a single field around the two vacua.
We will show in the following sections that the $N=2$ supersymmetry can be explicitly formulated on the lattice
(one of the transformations is essentially already given in (\ref{susymom1}) and (\ref{susymom2}) ), an invariant action can be constructed (including the mass and interaction term) and the continuum limit taken keeping exact supersymmetry at all stages. The doubling problem does not arise since the would be doublers are physical degrees of freedom in the same supermultiplet as the original field.

\section{The model}
\label{the-model} We briefly summarize the continuum version of
one dimensional $N=2$ supersymmetric model with two supersymmetry
charges $Q_1$ and $Q_2$~\cite{Cooper:1982dm}, whose matrix version
on the lattice was discussed in ~\cite{Arianos:2008ai}. Its
supersymmetry algebra is given by:
\begin{align}
Q_1^2=Q_2^2= P_t, ~~~&~~~\{Q_1,Q_2\}=0, \label{cont-susy-alg1} \\
[P_t,Q_1]=&[P_t,Q_2]=0, \label{cont-susy-alg2}
\end{align}
where $P_t$ is the generator of translations in the
one-dimensional space-time coordinate $t$\footnote{Unlike
reference ~\cite{Arianos:2008ai} we use here a Lorentzian metric.
The euclidean formulation of ~\cite{Arianos:2008ai} can be
obtained with a Wick rotation $t \rightarrow -i x$.}: \beq P_t =-i
\frac{\partial}{\partial t}. \label{hamilt} \eeq A superspace
representation of the algebra may be given in terms of two
Grassmann odd, real coordinates $\theta_1$ and $\theta_2$, namely:
\beq Q_1 = \frac{\partial}{\partial \theta_1} -i \theta_1
\frac{\partial}{\partial t},~~~~~~~~~~Q_2=\frac{\partial}{\partial
\theta_2} -i \theta_2 \frac{\partial}{\partial t}.
\label{susycharges} \eeq The field content of the theory is
described by a hermitian superfield $\Phi(t,\theta_1,\theta_2)$:
\beq \Phi(t,\theta_1,\theta_2) = \vph(t) + i \theta_1 \psi_1(t) +
i \theta_2 \psi_2(t) + i \theta_2 \theta_1 D(t),
\label{superfield} \eeq where $\psi_1$ and $\psi_2$ are Majorana
fermions. The supersymmetry transformations of the superfield
$\Phi$ are given by: \beq \delta_j \Phi = [ \eta_j Q_j, \Phi
]~~~~~~~~~~j=1,2, \label{susytrans} \eeq where $\eta_i$ are the
Grassmann odd parameters of the transformation. In terms of the
component fields eq. (\ref{susytrans}) reads:
\begin{align}
\delta_j \vph &= i \eta_j \psi_j, \\
\delta_j \psi_k&=  \delta_{j,k} \eta_j \partial_t \vph +
\epsilon_{jk} \eta_j D,
\\ \delta_j D &= i \epsilon_{jk} \eta_j \partial_t \psi_k.
\label{susycomp}
\end{align}
It is important to note that since the supersymmetry
transformations are defined in (\ref{susytrans}) as commutators,
supersymmetry transformations of superfields products obey Leibniz
rule: \beq \delta_i (\Phi_1 \Phi_2) = (\delta_i \Phi_1) \Phi_2 +
\Phi_1 (\delta_i \Phi_2). \label{LBr} \eeq In order to write a
supersymmetric action we need to introduce the super derivatives,
defined as \beq D_j = \frac{\partial}{\partial \theta_j} +i
\theta_j \frac{\partial}{\partial t}, \label{susyder} \eeq which
anticommute with the supersymmetry charges $Q_j$ and satisfy the
algebra: \beq D_j^2 = i \frac{\partial}{\partial
t},~~~~~~~~~~\{D_1,D_2\} = 0. \label{sderalg} \eeq The
supersymmetric action can then be defined in terms of the
superfield $\Phi$ as: \beq \int dt d\theta_1 d\theta_2 \left[
\frac{1}{2} D_2 \Phi D_1 \Phi + i V(\Phi) \right], \label{action}
\eeq where $V(\Phi)$ is a superpotential which may includes any
powers of superfields together with coupling constants. By
integrating over $\theta_1$ and $\theta_2$ in (\ref{action}) one
can obtain the action written in terms of the component fields. If
we take the super potential in the following form: \beq V(\Phi) =
\frac{1}{2} m \Phi^2 + \frac{1}{4} g \Phi^4, \label{potential}
\eeq we obtain the following action:
\begin{align}
S=& \int dt \{ \frac{1}{2} \left[  -(\partial_t \vph)^2 - D^2 +i
\psi_1 \partial_t \psi_1 +i \psi_2 \partial_t \psi_2 \right] \non
-& m(i \psi_1 \psi_2 + D \vph) - g ( 3 i \vph^2 \psi_1 \psi_2 + D
\vph^3 ) \}. \label{actioncomp}
\end{align}
As we can see, the general interaction terms in $\Phi^n$ are
of the forms of $\vph^{n-2} \psi_1 \psi_2 $ and $D \vph^{n-1}$.

It is convenient for later use to write  the SUSY transformations
in the Lorentzian metric and in the momentum representation. They
are given by:
\begin{align}
\delta_1 \vph(p) &= i \eta_1 \psi_1(p),    &\delta_2 \vph(p) &= i \eta_2 \psi_2(p), \nonumber  \\
\delta_1 \psi_1(p)&=  -i\eta_1 p \vph(p),  &\delta_2 \psi_1(p)&=  -\eta_2 D(p), \nonumber \\
\delta_1 \psi_2(p)&=  \eta_1 D(p) ,        &\delta_2 \psi_2(p)&=  -i \eta_2 p \vph(p) ,\nonumber \\
\delta_1 D(p) &=  \eta_1 p \psi_2(p),      &\delta_2 D(p) &=
-\eta_2 p \psi_1(p). \label{susycompLorMom}
\end{align}
Notice that $\delta_2$ is obtained from $\delta_1$ with the
discrete symmetry: $\psi_1 \rightarrow \psi_2$ an $\psi_2
\rightarrow -\psi_1$. Finally we write the  action in the momentum
representation:
\begin{align}
S=& \int dp \Bigl\{ \frac{1}{2} \left[  -p^2\vph(-p) \vph(p) -
D(-p)D(p) + i \psi_1(-p) p \psi_1(p) + i \psi_2(-p) p \psi_2(p)
\right] \non -& m(i \psi_1(-) \psi_2(p) + D(-p) \vph(p)) \Bigr\} \non -& g
\int dp_1 dp_2 dp_3 dp_4 ( 3 i \vph(p_1) \vph(p_2) \psi_1(p_3)
\psi_2(p_4) + D(p_1) \vph(p_2) \vph(p_3) \vph(p_4)  ) \delta(p_1 +
p_2 + p_3 + p_4). \label{actioncompLor}
\end{align}

\section{Supersymmetry transformations on the lattice}

\begin{figure}
 \hfil
 \includegraphics{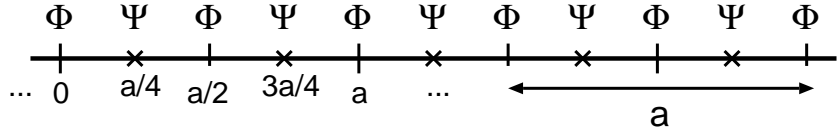}
  \caption{One dimensional lattice with the bosonic field $\Phi$ located on the
  points multiple of $\frac{a}{2}$, and the fermionic field $\Psi$ on points shifted by $\frac{a}{4}$.
 }
 \label{fig:1dlattice}
\end{figure}

According to the discussion of section \ref{basic-ideas} the formulation of the $D=1$, $N=2$ supersymmetric model on a lattice with spacing $\frac{a}{2}$ should involve on the lattice two fields, one bosonic and one fermionic. In fact the lattice consists of two sub lattices with lattice spacing $a$ invariant under translations and hence each degree of freedom on the lattice corresponds to two degrees of freedom in the continuum limit. Let us denote the bosonic ``superfield'' by $\Phi(x)$ with $x=\frac{n a}{2}$, and the fermionic ``superfield'' by $\Psi(x)$ with $x=\frac{n a}{2} + \frac{a}{4}$. The shift of $\frac{a}{4}$ in the fermionic superfield with respect to the bosonic one has been introduced to have symmetric finite differences in the supersymmetry transformations and implement hermiticity in a natural way as discussed in section \ref{basic-ideas}.
A picture of the lattice is given in fig.~\ref{fig:1dlattice}.

Let us proceed now to define the supersymmetry transformations on the lattice.
There are two supercharges in the $N=2$ model, whose algebra was given in (\ref{cont-susy-alg1}).
One of the  supersymmetry transformation, which we shall denote by $\delta_1$, was already formulated on the lattice in the context of the $N=1$ model and can be written as:
\bea
&&\delta_1 \Phi(x) = \frac{i \alpha}{2} \bigg[
\Psi(x+\frac{a}{4}) + \Psi(x-\frac{a}{4}) \bigg]~~~~~~~~x=\frac{n a}{2} \, ,\label{delta1b} \\
&& \delta_1 \Psi(x) = 2  \alpha \bigg[ \Phi(x+\frac{a}{4})
- \Phi(x-\frac{a}{4})  \bigg] ~~~~~~~~~x=\frac{n a}{2} + \frac{a}{4} \, . \label{delta1f} \eea
We assume here that $\Phi(x)$ and $\Psi(x)$ are dimensionless, so that no dependence on the lattice spacing $a$ appears at the r.h.s. of (\ref{delta1b}) and (\ref{delta1f}). Of course a rescaling of the fields with powers of $a$ will be needed to make contact with the fields of the continuum theory.
Let us introduce now the superfields in the momentum space defined as the Fourier transform of $\Phi(x)$ and $\Psi(x)$\footnote{For simplicity we shall denote fields in  momentum and coordinate representation with the same symbols. Arguments $x$,$y$, $z$ will always refer to coordinate representations, arguments $p$, $q$, $r$ to the momentum representation.}:
\beq
 \Phi(p) = \frac{1}{2} \sum_{x=\frac{n a}{2}} e^{ipx} \Phi(x), \, ~~~~~~~~~~~~~~~~~~\,
 \Psi(p) = \frac{1}{2} \sum_{x=\frac{n a}{2} + \frac{a}{4}} e^{ipx} \Psi(x). \, \label{ftrans}
\eeq The corresponding inverse transformations are:
 \beq
 \Phi(x) = a \int_0^{\frac{4\pi}{a}} \frac{d p}{2\pi} \Phi(p)e^{-ipx},\, ~~~~~~~~~~~~~~~~~~\,
\Psi(x) = a \int_0^{\frac{4\pi}{a}} \frac{d p}{2\pi}
\Psi(p)e^{-ipx}. \, \label{invtrans} \eeq From(\ref{ftrans}) it is
clear that $\Phi(p)$ and $\Psi(p)$ satisfy the following
periodicity conditions: \beq \Phi(p+\frac{4 \pi}{a} ) = \Phi(p),
\,~~~~~~~~~~~~~~~~~~\Psi(p+\frac{4 \pi}{a} ) = - \Psi(p).
\label{period} \eeq In momentum representation the supersymmetry
transformations (\ref{delta1b}) and (\ref{delta1f}) read: \bea
&&\delta_1 \Phi(p) =  i \cos \frac{a p}{4} \alpha \Psi(p) \, ,\label{delta1pb} \\
&& \delta_1 \Psi(p) = -4 i \sin \frac{a p}{4} \alpha  \Phi(p) \, .
\label{delta1pf} \eea The commutator of two supersymmetry
transformations $\delta_1$ with parameters $\alpha$ and $\beta$
defines an infinitesimal translation on the lattice. From
(\ref{delta1pb}) and (\ref{delta1pf}) one finds: \beq \delta_{1
\beta}\delta_{1 \alpha}F(p) - \delta_{1 \alpha}\delta_{1
\beta}F(p) = 4 \sin\frac{a p}{2} \alpha \beta F(p),  \label{inftr}
\eeq where $F(p)$ stands for either $\Phi(p)$ or $\Psi(p)$. In
coordinate space this is completely equivalent to (\ref{trs}), so
that an infinitesimal translation of parameter $\lambda$ on the
lattice is defined by; \beq F(x) \rightarrow F(x) + \lambda
\frac{F(x+\frac{a}{2}) - F(x-\frac{a}{2})}{a}, \label{inftrs} \eeq
which clearly reduces to $F(x)\rightarrow F(x + \lambda)$ in the
continuum limit. Translations defined in  (\ref{inftrs}) are
however conceptually different from the discrete translations on
the lattice that would be defined as: \beq F(x) \rightarrow F(x+a)
= F(x) + a \frac{F(x+a) -F(x)}{a} .   \label{dtrs} \eeq The
difference between the two definitions is even more apparent in
the momentum representation, where (\ref{dtrs}) is simply
$F(p)\rightarrow e^{iap}F(p)$ and applied to a product of fields
leads to the standard form of momentum conservation. Invariance
under (\ref{inftrs}) instead leads to a non local conservation law
where $p$ is replaced by $\sin\frac{a p}{2}$, namely , for a
product of fields of momenta $p_1$,$p_2$,...,$p_n$: \beq
\sin\frac{a p_1}{2}+\sin\frac{a p_2}{2}+ \cdots+\sin\frac{a
p_n}{2}=0 .\label{sincons} \eeq This conservation law on the lattice
was first pointed out by Dondi and Nicolai~\cite{D-N}.
The implications of this conservation law, in particular with respect to the validity of the Leibniz rule and the relation of the present approach to the link approach will be discussed in section 7. In the
continuum limit ($a p_i \ll1$)
(\ref{sincons}) reduces to the standard momentum conservation
law and locality is restored. The conservation law (\ref{sincons})
is not affected if any momentum $p_i$ in it is replaced by
$\frac{2 \pi}{a} - p_i$ due to the invariance of the sine. In view
of last section's discussion the interpretation is clear: in the
continuum limit ( $ a p \ll 1$) $F(p) $ and $F(\frac{2 \pi}{a} -
p)$ represent fluctuations of momentum $p$ respectively around the
vacuum of  momentum zero and $\frac{2 \pi}{a}$ on the lattice. So
the symmetry $p \rightarrow \frac{2 \pi}{a} - p$ amounts to
exchanging the two vacua keeping the physical momentum unchanged
and will play an important role in supersymmetry transformations.

We now want to match the superfields $\Phi(p)$ and $\Psi(p)$ appearing in the supersymmetry transformations (\ref{delta1pb}) and (\ref{delta1pf}) with the component fields of the $N=2$ $D=1$ supersymmetry  described in the previous section. Working in the momentum representation we shall associate $\varphi$ and $D$ with the fluctuations of $\Phi$ respectively around $0$ and $\frac{2 \pi}{a}$ and similarly $\psi_1$ and $\psi_2$ with the fluctuations of $\Psi$.
More specifically we assume the following correspondence:
\bea
&& \Phi(p) = a^{-\frac{3}{2}} \varphi(p), \label{correspondencei}   \\
&& \Psi(p) = a^{-1} \psi_1(p), \label{corr1}\\
&& \Psi(\frac{2 \pi}{a} - p) = i a^{-1} \psi_2(p) ,\label{corr2}\\
&& \Phi(\frac{2 \pi}{a} - p) = -\frac{a^{-\frac{1}{2}}}{4} D(p),
\label{correspondence} \eea where $p$ is  restricted in
(\ref{correspondencei}-\ref{correspondence}) to the interval
$(-\frac{\pi}{a}, \frac{\pi}{a})$, which is also the range of
definition of the component fields  (although with a possible
discontinuity at $p=\pm \frac{\pi}{a}$) which corresponds to a
lattice of spacing $a$.  A rescaling of the fields with powers of
$a$ has been also introduced in
(\ref{correspondencei}-\ref{correspondence}) to account for the
dimensionality of the component fields in momentum representation
(remember that $\Phi$ and $\Psi$ were defined to be
dimensionless). A similar rescaling will be assumed for the
supersymmetry parameter $\alpha$: \beq \alpha = a^\frac{1}{2} \eta.
\label{alpha} \eeq By inserting
(\ref{correspondencei}-\ref{correspondence}) and (\ref{alpha})
into the supersymmetry transformations (\ref{delta1pb}) and
(\ref{delta1pf}) we obtain for the component fields the following
transformations: \bea
 \delta_1 \varphi(p) = i \cos\frac{ap}{4} \eta \psi_1(p) & \underset{ap \ll 1}{\longrightarrow} &  i \eta \psi_1(p),  \label{susy1comp1}\\
 \delta_1 \psi_1(p) = -i \frac{4}{a} \sin\frac{ap}{4} \eta \varphi(p) &   \underset{ap \ll 1}{\longrightarrow} & -i p \eta \varphi(p),\\
 \delta_1 \psi_2(p) = \cos\frac{ap}{4} \eta D(p) &  \underset{ap \ll 1}{\longrightarrow} & \eta D(p),, \\
 \delta_1 D(p) = \frac{4}{a} \sin\frac{ap}{4} \eta \psi_2(p) &  \underset{ap \ll 1}{\longrightarrow} & p \eta \psi_2(p) . \label{susy1comp2}
 \eea
  In the continuum limit $ a p \ll 1$  the above supersymmetry transformations on the lattice reproduce exactly the ones
  in the continuum (\ref{susycompLorMom}) given in Section 3.
%\ref{}.
  As already mentioned after eq. (\ref{correspondence}), the momentum $p$ appearing as the argument of the component fields in (\ref{correspondencei}-\ref{correspondence}) and in (\ref{susy1comp1}-\ref{susy1comp2}) is restricted to the interval $(-\frac{\pi}{a}, \frac{\pi}{a})$. So we could introduce a lattice of spacing $a$ and coordinates $\tilde{x}= n a$ and define the component fields $\varphi(\tilde{x})$, etc. on such lattice by taking the Fourier transform on the $\frac{2\pi}{a}$ interval of $\varphi(p)$, etc. and finally write the supersymmetry transformations (\ref{susy1comp1}-\ref{susy1comp2}) of the component fields in the $\tilde{x}$ coordinate representation. However the trigonometric functions at the r.h.s. of (\ref{susy1comp1}-\ref{susy1comp2}) are not periodic of period $\frac{2\pi}{a}$, and as a result the supersymmetry transformation are non-local in the $\tilde{x}$ coordinate representation, implying that the natural representation for the supersymmetry is on the lattice with $\frac{a}{2}$ spacing.

  We have now to identify the second supersymmetry transformation $\delta_2$. In the continuum $\delta_2$ is obtained from $\delta_1$ by replacing everywhere $\psi_1(p)$ with $\psi_2(p)$, and $\psi_2(p)$ with $-\psi_1(p)$. It is easy to see from (\ref{correspondencei}-\ref{correspondence}) that this corresponds on the lattice to the replacement:
  \beq
  \Psi(p) \longrightarrow -i \Psi(\frac{2\pi}{a}-p).  \label{replacement}
  \eeq
  By performing this replacement on the supersymmetry transformations (\ref{delta1pb}) and (\ref{delta1pf})
  one obtains the expression for $\delta_2$:
  \bea
&&\delta_2 \Phi(p) =   \cos \frac{a p}{4} \alpha \Psi(\frac{2\pi}{a}-p) \, ,\label{delta2pb} \\
&& \delta_2 \Psi(\frac{2\pi}{a}-p) = 4  \sin \frac{a p}{4} \alpha  \Phi(p) \, . \label{delta2pf} \eea
The supersymmetry transformation $\delta_2$, defined by (\ref{delta2pb}) and (\ref{delta2pf}), satisfies together with $\delta_1$ an $N=2$ supersymmetry algebra. It is easy to check in fact that the commutator of two $\delta_2$ transformations gives an infinitesimal translation ( namely eq. (\ref{inftr}) holds also for $\delta_2$) and that the commutator of a $\delta_1$ and $\delta_2$ transformation vanishes, namely:
  \beq
\delta_{1 \beta}\delta_{2 \alpha}F(p) - \delta_{2 \alpha}\delta_{1 \beta}F(p) = 0.  \label{inftrm}
  \eeq
  In terms of the component fields, and in the continuum limit, the explicit expression for the $\delta_2$ transformation can be obtained from (\ref{delta2pb}) and (\ref{delta2pf}) by using eq.s (\ref{correspondencei}-\ref{correspondence}):
  \bea
 \delta_2 \varphi(p) = i \cos\frac{ap}{4} \eta \psi_2(p) & \underset{ap \ll 1}{\longrightarrow} &  i \eta \psi_2(p),  \label{susy2comp1}\\
 \delta_2 \psi_2(p) = -i \frac{4}{a} \sin\frac{ap}{4} \eta \varphi(p) &   \underset{ap \ll 1}{\longrightarrow} & -i p \eta \varphi(p),\\
 \delta_2 \psi_1(p) = -\cos\frac{ap}{4} \eta D(p) &  \underset{ap \ll 1}{\longrightarrow} & -\eta D(p), \\
 \delta_2 D(p) = -\frac{4}{a} \sin\frac{ap}{4} \eta \psi_1(p) &  \underset{ap \ll 1}{\longrightarrow} &- p \eta \psi_1(p).  \label{susy2comp2}
 \eea
 As for $\delta_1$ in the  limit $ap \ll 1$ the above transformation coincides, in the momentum space representation, with the one generated by $Q_2$ in the continuum theory.

 The coordinate representation of $\delta_2$ can be obtained directly from (\ref{susy2comp1}-\ref{susy2comp2}) by Fourier transform, or from (\ref{delta1b}-\ref{delta1f}) by performing the following substitution:
\beq \Psi(x) \longrightarrow (-1)^n
\Psi(-x)~~~~~~~~~~~~x=\frac{na}{2}-\frac{a}{4},~~~~~~~~~~~~~
\label{replacementx} \eeq which is the same as (\ref{replacement})
in the coordinate representation. Either way the result is: \bea
&&\delta_2 \Phi(x) = \frac{i \alpha}{2} (-1)^n\bigg[
\Psi(-x+\frac{a}{4}) - \Psi(-x-\frac{a}{4}) \bigg]~~~~~~~~x=\frac{n a}{2} \, ,\label{delta2b} \\
&& \delta_2 \Psi(x) = 2  \alpha (-1)^n\bigg[ \Phi(-x+\frac{a}{4})
- \Phi(-x-\frac{a}{4})  \bigg] ~~~~~~~~~x=\frac{n a}{2} +
\frac{a}{4} \, . \label{delta2f} \eea

It is clear from (\ref{delta2b}) and (\ref{delta2f}) that the
supersymmetry transformation $\delta_2$ is local in the coordinate
representation only modulo the  reflection $x \rightarrow -x$.
This was already implicit in the correspondence
(\ref{correspondencei}-\ref{correspondence}) between the lattice
fields and the ones of the continuum theory. In fact it is clear
from (\ref{correspondencei}-\ref{correspondence}) that while for
instance $\varphi(x)$ is associated to the fluctuations of
$\Phi(x)$ around the constant configuration ($p=0$), the
fluctuations of $\Phi(x)$ around the constant configuration with
alternating sign ($p=\frac{2\pi}{a}$) correspond in the continuum
to $D(-x)$. For fermions this parity change leads to a physical
meaning. Since $\psi_2(p) \leftrightarrow \psi_2(x)$ is defined as
a species doubler of $\psi_1(p) \leftrightarrow \psi_1(x)$, the
chirality of $\psi_2$ is the opposite of $\psi_1$. However by the
change of $p \rightarrow \frac{2\pi}{a}-p $ equivalently $x
\rightarrow -x$, chirality of $\psi_1$ and $\psi_2$ are adjusted
to be the same. Thus this bi local nature in the coordinate space
may be transfered to a local interpretation.

\section{Supersymmetric invariant action}

In order to construct a lattice action invariant under the two supersymmetry transformations $\delta_1$ and $\delta_2$ defined in the previous section we consider first the invariance under translations, which follows from supersymmetry, and it is expressed by the sine conservation law given in (\ref{sincons}).

Any vertex of a supersymmetric invariant theory will have to include a delta function enforcing the conservation law (\ref{sincons}). Unlike the standard momentum conservation this conservation law does not lead to a local action
in coordinate space, and in fact it makes it impossible to write the action in coordinate space without
using transcendental function (more specifically Bessel functions). For this reason
we shall first formulate the action  in the momentum representation. We provide a full treatment of coordinate prescription later in section 7.
There is however an important exception to this, namely the case $n=2$, that is the kinetic term and the mass term. In fact for $n=2$ the conservation law (\ref{sincons}) has two solutions:
\beq
p_1+ p_2=0~~~~~~~~~~~~~~~~~~({\rm mod}~~~~ \frac{4 \pi}{a})  \label{mcon1}
\eeq
and
\beq
p_1 - p_2 = \frac{2 \pi}{a}~~~~~~~~~~~~~~~~~~({\rm mod}~~~~ \frac{4\pi}{a}) \label{mcon2}
\eeq
and the delta function of momentum conservation does not need in this case to include any sine function. The two point term in the action with the momentum conservation (\ref{mcon1}) describes, as we shall see, the kinetic term and is local when expressed in the coordinate representation. The mass term will be expressed instead by a term with the conservation law (\ref{mcon2}), involving in coordinate space a coupling between fields in $x$ and $-x$, in agreement with the discussion at the end of the previous section.
Let us introduce now a supersymmetric action on the lattice. All  terms of this action have the same structure, which for an  $n$-point term is the following:
\begin{align}
 S^{(n)}
 &=g_0^{(n)}a^n \frac{4}{n!} \int_{-\frac{\pi}{a}}^{\frac{3\pi}{a}} \frac{d p_1}{2\pi} \cdots \frac{d p_n}{2\pi}
   2\pi\delta\left(\sum_{i=1}^n\sin\frac{a p_i}{2}\right)
   \nonumber\\
 &\qquad\times
  G(p_1,p_2,\cdots,p_n)\Bigl[
  2 \sin^2\frac{a p_1}{4} ~\Phi(p_1)\Phi(p_2)\cdots\Phi(p_n) + \label{kterm}\\
 & + \frac{n-1}{4}~ \sin\frac{a(p_1-p_2)}{4}~ \Psi(p_1)\Psi(p_2)\Phi(p_3)\cdots\Phi(p_n)\Bigr].
  \nonumber
\end{align}
A direct check shows that $S^{(n)}$ is invariant under the
supersymmetry transformation $\delta_1$ defined in
(\ref{delta1b},\ref{delta1f}) as well as under the replacement
(\ref{replacement}), which in turn implies the invariance under
$\delta_2$ \emph{provided} the otherwise arbitrary function $G$
satisfies the following properties: \emph{i}) it is symmetric
under permutations of the momenta $p_i$, \emph{ii}) it is
periodic with period $\frac{4\pi}{a}$ in all the momenta and \emph{iii})  it
is invariant when an \emph{even} number of $p_i$ is replaced by
$ \frac{2\pi}{a} - p_i$. An example of function satisfying the above
requirements is:
\beq C(p) = \prod_{i=1}^n
~\cos\frac{a p_i}{2}.     \label{cosfunc} \eeq
For all interaction terms ($n > 2$) we will take $G(p) = C(p)$.
In fact, thanks to the cosine factors the function $C(p)$  vanishes if any of the momenta $p_i$ is equal to $\pm \frac{\pi}{a}$, so a factor $C(p)$ is needed to cancel the singularities arising at $p_i=\pm \frac{\pi}{a}$ from the integration
 volume as a consequence of the delta function.
%  Besides, $C(p)\sim 1$ (resp. $C(p)\sim -1$) if an
%even (resp. odd) number of momenta are close to the vacuum at
%$p= \frac{2\pi}{a}$ and the remaining close to the vacuum at $p=0$. Thus
%$C_{\pm}(p)=1/2( 1  \pm C(p))$ are in the continuum limit projectors onto terms %with definite chirality.
Later in section seven we find out another natural reason why this factor
(\ref{cosfunc}) comes out.
All terms in (\ref{kterm}) are periodic with period $\frac{4 \pi}{a}$ in the momenta, and the momentum integration is over a whole period: the specific choice here (from $\frac{-\pi}{a}$ to $\frac{3 \pi}{a}$) is for future convenience.
The integrand can be made explicitly symmetric with respect to permutations of the momenta, so the factor $2 \sin^2\frac{a p_1}{4}$ in front of
the bosonic part can be replaced by
 \beq
2 \sin^2\frac{a p_1}{4} \rightarrow  1 - \frac{1}{n} \sum_1^n \cos\frac{a p_i}{2}.
\label{smc}
\eeq
It should be noticed also that, although we kept the dependence on the lattice spacing $a$ explicit, this could be completely absorbed in the definition of the momenta, by introducing a-dimensional momenta $\tilde{p}_i = a p_i$.

\subsection{Kinetic term and Mass term}

The case $n=2$ is special because only in that case the sine conservation law
splits into the two separate conservation laws  (\ref{mcon1}) and (\ref{mcon2}) which are linear in the momenta. The delta function in (\ref{kterm}) can then be replaced by the sum, with arbitrary coefficients, of the delta functions enforcing
(\ref{mcon1}) and (\ref{mcon2}).
This amounts ( for $n=2$ only) to perform in (\ref{kterm}) the following replacement:
\beq
\frac{a}{2} g_0^{(2)}G(p_1,p_2) \delta(\sin\frac{a p_1}{2}+\sin\frac{a p_2}{2}) \longrightarrow \delta(p_1+p_2) + m_0 \delta(p_1-p_2-\frac{2\pi}{a}),     \label{deltas}
\eeq
where $m_0$ is a free parameter. The delta functions at the r.h.s. of (\ref{deltas}) do not give rise to any singularity at $p_i = \pm \frac{\pi}{a}$ so no factor $C(p)$ is required in this case\footnote{Of course it would be possible to treat the $n=2$ case in the same way as the interaction terms, keeping the sine delta function with the factor $C(p)$ in front. However this would fix the relative coefficient  of the mass term and of the kinetic term. The smoothness of the continuum limit would then be spoiled since, as we shall see,  $m_0$ need to scale with $a$ for such limit to be smooth.}
We are going to show here that the first delta in (\ref{deltas}) generates  the supersymmetric kinetic term, the second delta  the supersymmetric mass term.
By inserting the r.h.s. of (\ref{deltas}) into (\ref{kterm}) and performing one momentum integration we obtain:
\beq
S^{(2)} = S_{kin} + S_{mass} ,              \label{s2}
\eeq
with
\beq
S_{kin} = 4a \int_{-\frac{\pi}{a}}^{\frac{3\pi}{a}} \frac{dp}{2\pi} \left[ 2
\sin^2\frac{ap}{4} \Phi(-p)\Phi(p) - \frac{1}{4} \sin \frac{ap}{2} \Psi(-p)\Psi(p)\right] ,          \label{skin}
\eeq
and
\beq
S_{mass} = 4a m_0 \int_{-\frac{\pi}{a}}^{\frac{3\pi}{a}} \frac{dp}{2\pi} \left[ \Phi(p+\frac{2\pi}{a})\Phi(p) + \frac{1}{4} \Psi(p+\frac{2\pi}{a})\Psi(p)\right].
\label{smass}
\eeq
One could regard the kinetic term (\ref{skin}) as the kinetic term of a lattice theory with lattice spacing $a' = \frac{a}{2}$ and just one bosonic and one fermionic degree of freedom. The invariance under the $\delta_1$ supersymmetry transformations (\ref{delta1pb},\ref{delta1pf}) would be described as an $N=1$ supersymmetry. However, as shown by (\ref{skin}), the fermion would have a doubler at $p= \frac{\pi}{a'}$, as expected.
Our interpretation is different. Since supersymmetry transformations (\ref{delta1f}) are related to shifts of $\frac{a}{2}$, we consider $a$ as the fundamental lattice spacing and the $\frac{a}{2}$ spacing as the signal that we are describing with the same lattice field two distinct degrees of freedom in the continuum: hence the fermion and its  doubler are interpreted as partners in an $N=2$ supersymmetry generated by $\delta_1$ and $\delta_2$, the latter given by (\ref{delta2pb},\ref{delta2pf}).
In order to separate the degrees of freedom let us split the integration region in (\ref{skin}) and (\ref{smass}) into $(-\frac{\pi}{a},\frac{\pi}{a})$ and $(\frac{\pi}{a},\frac{3 \pi}{a})$. In the first interval we use the correspondence (\ref{correspondencei},\ref{corr1}), in the second the correspondence (\ref{corr2},\ref{correspondence}) and find:
\bea
S_{kin} = &&\int_{-\frac{\pi}{a}}^{\frac{\pi}{a}} \frac{d p}{2\pi} \Bigl[
\frac{4}{a^2} (1-\cos\frac{a p}{2}) \varphi(-p)\varphi(p) +\frac{1}{4}(1+\cos\frac{a p}{2}) D(-p)D(p)-\nonumber \\ &&- \frac{1}{a} \sin\frac{a p}{2}\psi_1(-p)\psi_1(p)-\frac{1}{a} \sin\frac{a p}{2}\psi_2(-p)\psi_2(p)  \Bigr].\label{kincomp}
\eea
Similarly for the mass term we get:
\beq
S_{mass} = 2 m \int_{-\frac{\pi}{a}}^{\frac{\pi}{a}} \frac{d p}{2\pi} \left[ -\varphi(-p)D(p) -i \psi_1(-p)\psi_2(p) \right], \label{masscomp}
\eeq
where $m$ is now the physical mass: $m = \frac{m_0}{a}$. Thanks to the rescaling all fields in (\ref{kincomp}) and (\ref{masscomp}) have the correct canonical dimension, and the continuum limit is smooth.
The component fields $\varphi(p)$,$D(p)$,$\psi_1(p)$ and $\psi_2(p)$ are defined for $p$ in the interval $(-\frac{\pi}{a},\frac{\pi}{a})$. This is the Brillouin zone corresponding to a lattice of spacing $a$, so we could define a lattice with coordinates $\tilde{x}= n a$ and the component fields on it as the Fourier transforms of the momentum space components. For instance we could define:
\beq
\varphi(\tilde{x}) =  \int_{-\frac{\pi}{a}}^{\frac{\pi}{a}}\frac{d p}{2\pi}  \varphi(p) e^{-ip\tilde{x}}.
\label{tildex}
\eeq
However the action written in the coordinate $\tilde{x}$ space is non-local, since the finite difference operators appearing in (\ref{kincomp}) are periodic with period $\frac{4 \pi}{a}$ and not $\frac{2\pi}{a}$ that would be needed for a local expression on a lattice with spacing $a$.

Instead it is possible, using (\ref{ftrans}), to write (\ref{skin}) in the coordinate space $x$ with lattice spacing $\frac{a}{2}$:
\beq
S_{kin} =  \sum_{x=n\frac{a}{2}} \left[ \Phi(x) \left( 2 \Phi(x) - \Phi(x+\frac{a}{2})-\Phi(x-\frac{a}{2}) \right) + \frac{i}{2} \Psi(x+\frac{a}{4})\Psi(x-\frac{a}{4}) \right]. \label{coordskin}
\eeq
In the bosonic part of the action the equivalent of a second finite difference appears. In fact if we define the finite difference on the lattice of spacing $\frac{a}{2}$ as
\beq
\partial_{\pm} F(x) = F(x\pm \frac{a}{2}) - F(x),    \label{fdiff}
\eeq the kinetic term can be rewritten as: \beq S_{kin} =
 \sum_{x=n\frac{a}{2}} \left[ \Phi(x)
~\partial_{-}\partial_{+} \Phi(x)+ \frac{i}{2}
\Psi(x+\frac{a}{4})~\partial_{-}\Psi( x+\frac{a}{4})\right].
\label{coordskin2} \eeq The coordinate representation for the mass
term (\ref{smass}) reveals some new features, namely a coupling
between fields in $x$ and $-x$. In fact, by using again
(\ref{ftrans}), one finds: \beq S_{mass}= \frac{m_0}{2}
\sum_{x=n\frac{a}{2}} (-1)^{\frac{2x}{a}} \left[ \Phi(-x)\Phi(x) +
\frac{i}{4} \Psi(-x-\frac{a}{4})\Psi(x+\frac{a}{4}) \right].
\label{coordsmass} \eeq The bi local structure of
(\ref{coordsmass}) shows that the extended lattice  with spacing
$\frac{a}{2}$ has not a straightforward relation to the coordinate
space in the continuum limit. This is related to the fact that
while the fluctuations of $\Phi(x)$ (resp. $\Psi(x)$) around a
constant field configuration are associated to the component field
$\varphi(x)$ (resp. $\psi_1(x)$), its fluctuations around
$(-1)^{\frac{2 x}{a}}$ are associated to $D(-x)$ (resp.
$\psi_2(-x)$). In other words the way the two bosonic (resp. fermionic)
components of the superfield are embedded in a single bosonic
(resp. fermionic) field on the extended lattice  is non trivial
and exhibits a bi local structure. Although the extended lattice is
not a discrete representation of superspace (bosonic and fermionic
fields have to be introduced separately on it) it carries some
information about the superspace structure and as such it does not
simply map onto the coordinate space in the continuum limit.

From (\ref{kincomp}) and (\ref{masscomp})one can then easily
derive the free propagators. For this purpose it is convenient to
introduce the following notations: \beq \hat{p} = \frac{2}{a}
\sin\frac{ap}{2}, \label{hp} \eeq and \beq c(\hp)=\sqrt{1-\frac{a^2
\hp^2}{4}}. \label{cq}\eeq Moreover, for each component field
$f(p)$ we define \beq f'(\hp) = f(p(\hp)). \label{fprime} \eeq
With these notations the two point bosonic correlation function
can be written as:
\begin{align}
 \langle \varphi'(\hp) \varphi'(-\hp) \rangle
  &=\frac{c(\hp)}{\hp^2 -4m^2}\frac{1}{2}\left(1+c(\hp)\right),
\label{eq:propagatorp-phiphi}\\
 \langle D'(\hp) D'(-\hp) \rangle
  &=\frac{c(\hp)}{\hp^2 -4m^2}\frac{8}{a^2}\left(1-c(\hp)\right),\\
 \langle \varphi'(\hp) D'(-\hp) \rangle
  &=\frac{c(\hp)}{\hp^2 -4m^2} 2m,
\end{align}
whereas for the fermionic ones we have:
\begin{align}
 \langle \psi_1'(\hp)\psi_1'(-\hp) \rangle
  =\langle \psi_2'(\hp)\psi_2'(-\hp) \rangle
  &= - \frac{c(\hp)}{\hp^2-4m^2} a\hp, \\
 \langle \psi_1'(\hp)\psi_2'(-\hp) \rangle
  = - \langle \psi_2'(\hp)\psi_1'(-\hp) \rangle
  &= \frac{c(\hp)}{\hp^2-4m^2} 2im.
 \label{eq:propagatorp-psi1psi2}
\end{align}

\subsection{Interaction terms}

The interaction terms are obtained from the general invariant
expression (\ref{kterm}) with $n > 2$. We shall choose the
arbitrary function $G$ to be equal to the function $C(p)$ defined
in (\ref{cosfunc}). This is needed to cancel the divergences
occurring in the integration volume at $p_i = \pm \frac{\pi}{a}$
due to the delta function. With this choice the $n$-point
interaction term reads:
\begin{align}
 S^{(n)}
 &=g_0^{(n)}a^n \frac{4}{n!}\int_{-\frac{\pi}{a}}^{\frac{3\pi}{a}} \frac{d p_1}{2\pi} \cdots \frac{d p_n}{2\pi}
   2\pi\delta\left(\sum_{i=1}^n\sin\frac{a p_i}{2}\right)
   \nonumber\\
 &\qquad\times
  \left(\prod_{i=1}^n \cos\frac{ap_i}{2}\right)\Bigl[
  2 \sin^2\frac{a p_1}{4} ~\Phi(p_1)\Phi(p_2)\cdots\Phi(p_n) + \label{nterm}\\
 & + \frac{n-1}{4}~ \sin\frac{a(p_1-p_2)}{4}~ \Psi(p_1)\Psi(p_2)\Phi(p_3)\cdots\Phi(p_n)\Bigr].
  \nonumber
\end{align}
Unlike the case of $S^{(2)}$ for $n \geq 3$ the delta function of
momentum conservation is not linear in $p_i$, hence the coordinate
representation for $S^{(n)}$ cannot be given in terms of elementary
functions and it is non-local.

When expressed in terms of the
component fields (\ref{nterm}) contains many terms, as each field
in (\ref{nterm}) can correspond to  different components of the
superfield depending on the value of its momentum. These terms
however have different powers of the lattice spacing $a$ according
to the rescaling properties given in
(\ref{correspondencei}--\ref{correspondence}). We have therefore to
select the terms that are leading in the continuum limit. In the
bosonic sector $D$ scales with an extra power of $a$ with respect
to $\varphi$, so that the leading term seems to be obtained by replacing
$\Phi(p_i)$ with $a^{-\frac{3}{2}} \varphi(p_i)$ and restricting
$p_i$ between $-\frac{\pi}{a}$ and $\frac{\pi}{a}$.
This term, however, is not the leading term, because of a factor
with the momentum labelled $p_1$ in the bosonic part of the action.
The $\sin^2$ factor multiplying $\Phi(p_1)$ is of order $a^2$ at
$p_1\simeq 0$ and of order $1$ at $p_1 \simeq \frac{2\pi}{a}$, so
that the latter vacuum becomes dominant and $\Phi(p_1)$ should be
identified with $D$.  That is, the leading term in the bosonic sector
is $D\varphi^{n-1}$.
%When expressed in terms of the
%component fields (\ref{nterm}) contains many terms, as each field
%in (\ref{nterm})can correspond to  different components of the
%superfield depending on the value of its momentum. These terms
%however have different powers of the lattice spacing $a$ according
%to the rescaling properties given in
%(\ref{correspondencei}-\ref{correspondence}). We have therefore to
%select the terms that are leading in the continuum limit. In the
%bosonic sector $D$ scales with an extra power of $a$ with respect
%to $\varphi$, so that the leading term is obtained by replacing
%$\Phi(p_i)$ with $a^{-\frac{3}{2}} \varphi(p_i)$ and restricting
%$p_i$ between $-\frac{\pi}{a}$ and $\frac{\pi}{a}$. There is
%however an exception which is the factor with the momentum
%labeled $p_1$ in the bosonic part of the action. In that case the
%$\sin^2$ factor multiplying $\Phi(p_1)$ is of order $a^2$ at
%$p_1\simeq 0$ and of order $1$ at $p_1 \simeq \frac{2\pi}{a}$, so
%that the latter vacuum becomes dominant and $\Phi(p_1)$ should be
%identified with $D$.
In the fermionic part of the action $\Psi(p)$
scales in the same  way at $p=0$ and $p=\frac{2\pi}{a}$, but
$p_1-p_2$ has to be $\frac{2\pi}{a}$ and not zero to avoid an
extra factor $a$ coming from the $\sin$ factor. So $\Psi(p_1)$ and
$\Psi(p_2)$ must correspond one to $\psi_1$ and one to $\psi_2$.
By carefully counting the powers of $a$, one finds that in order
to have a finite and non vanishing continuum limit for the leading
term of (\ref{nterm}) the physical coupling constant $g^{(n)}$
must be defined as: \beq g^{(n)}= a^{-\frac{n}{2}} g_0^{n}
\label{coupling} \eeq

With this normalization the leading term in the interaction term
becomes:
\begin{align}
 S^{(n)}
 &=\frac{g^{(n)}}{n!} \int_{-\frac{\pi}{a}}^{\frac{\pi}{a}} \frac{d p_1}{2\pi} \cdots \frac{d p_n}{2\pi}
   2\pi\delta\left(\frac{2}{a} \sum_{i=1}^n\sin\frac{a p_i}{2}\right)
   \nonumber\\
 &\qquad\times
  \left(\prod_{i=1}^n \cos\frac{ap_i}{2}\right)\Bigl[
  - \cos^2\frac{a p_1}{4} ~D(p_1)\varphi(p_2)\cdots\varphi(p_n) + \label{ntermcomp}\\
 & + (n-1)~ \cos\frac{a(p_1+p_2)}{4}~
 \psi_2(p_1)\psi_1(p_2)\varphi(p_3)\cdots\varphi(p_n)\Bigr]+ O(a),
  \nonumber
\end{align}
where the terms of order $a$ or higher are included in $O(a)$.
The leading term corresponds to the usual $\Phi^n$ interaction:
\beq S_i^{(n)} = \int dx d^2\theta \Phi^n. \label{supint} \eeq
The $O(a)$ terms are of two types: some contain higher powers of
$D(p)$,  namely terms that do not appear in the continuum in any
superfield action for dimensional reasons, but  are needed on
the lattice for exact supersymmetry. Then there are terms where all momenta
are fluctuating around the  $p=0$ vacuum and have the same structure
as the kinetic term. They correspond in the continuum to
derivative interactions given in terms of the superfield $\Phi$ by
 \beq S_k^{(n)} = \int dx
d^2\theta ~\Phi^{n-2} D_1 \Phi ~D_2\Phi. \label{supkin} \eeq
These derivative interaction terms are sub leading (of order $a$) with
the  choice of the function $G(p)$ given above, namely
$G(p)=C(p)$. However with a different choice of $G(p)$ they can be the leading
terms in the continuum limit.
 For instance, if we choose $G(p)=\frac{1}{a} C(p)(1+C(p))$, kinetic-like terms in (\ref{kterm}) with $0$, or $2$ momenta around the vacuum at $\frac{2\pi}{a}$ would be of order $1$  in the continuum limit\footnote{The reason is that $C(p)$ is $1$ when an even number of momenta are equal to $\frac{2\pi}{a}$ and the remaining are zero, it is $-1$ if the number of momenta equal to $\frac{2\pi}{a}$ is odd.} which would contain  derivative interactions of the form (\ref{supkin}).
  It is rather surprising that
the same action on the lattice, namely the one given in
(\ref{kterm}), can give origin to terms which are entirely
different when written in the  superfield formalism. This seem to indicate that
some deeper understanding of supersymmetry may be achieved by the
present approach.

\section{ Continuum limit and Ward identities}

One of the key features of the present approach is that momentum
conservation is replaced by the $\sin$ conservation law
(\ref{sincons}). This means that invariance under finite
translations is violated on the lattice. It is then crucial that
translational invariance is recovered in the continuum limit. This
is not obvious and it requires the analysis of the UV properties
of the theory when the continuum limit is taken. Recovery of
translational invariance is the subject of the first part of this
section. In the second part we check explicitly at one loop level
that supersymmetric Ward-Takahashi identities are preserved in the
continuum limit.

\subsection{ Continuum limit and translational invariance}

As a preliminary step towards a proof that invariance under finite translations
is recovered in the continuum limit, we proceed to analyze such a limit in the ultraviolet region.

The lattice theory described in the previous section  in terms of
the fields $\Phi$ and $\Psi$ is free of ultraviolet divergences.
In fact everything in that theory can be written in terms of the
dimensionless  momentum variables $\tilde{p}_i= a p_i$, which are
 angular variables with periodicity $ 4\pi$. Momentum
integrations reduce to integrations over trigonometric functions
of $\tilde{p}_i$, and ultraviolet divergences never appear. All
correlation functions of $\Phi$s and $\Psi$s integrations are
therefore finite. This however is not enough to ensure that the
continuum limit is smooth and that ultraviolet divergences do not
appear in the limiting process. The continuum limit in fact
involves  a rescaling of fields with powers of $a$, which is
singular in the $a \rightarrow 0$ limit. At the same time the
continuum limit, being a limit where $a \rightarrow 0$ keeping the
physical momentum fixed, corresponds to a situation where all
external momenta $\tilde{p}_i$ are in the neighborhood of one of
the vacua, namely at $\tilde{p}_i=0$ or $\tilde{p}_i=2 \pi$. The
limit being a singular one, the ultraviolet behavior has to be
checked explicitly.

Let us consider then the action written in
terms of the rescaled  component fields. The structure of its
interaction terms (neglecting momentum integrations and delta
functions) is the following:
\begin{align}
 S_{\rm B}^{(n)} &\sim \frac{g^{(n)}}{a} (aD)^k \varphi^{n-k}, \label{intstrb}\\
 S_{\rm F}^{(n)} &\sim \frac{g^{(n)}}{a} a \psi_i \psi_j (aD)^k
 \varphi^{n-k-2}.\label{intstrf}
\end{align}
Therefore it is convenient to introduce for the internal lines in
loop integrations the rescaled fields $D_{\rm int}= aD$ and
$\psi_{i, {\rm int}}=a^{\frac{1}{2}} \psi_i$. In this way the
effective coupling in the perturbative expansion is $g^{(n)}/a$
and each vertex contributes with a factor $\frac{1}{a}$ in the
ultraviolet region. Next we consider the UV behavior (including
momentum integration in the variable
$\hat{p}=\frac{2}{a}\sin\frac{ap}{2}$) of the propagators:
\begin{align}
 \langle \varphi \varphi \rangle
   \sim \langle D_{\rm int} D_{\rm int} \rangle
  &\sim \int^{2/a} d\hat{p} \frac{1}{\hat{p}^2 -4m^2} \sim a, \\
 \langle \varphi D_{\rm int} \rangle
  &\sim \int^{2/a} d\hat{p} \frac{am}{\hat{p}^2 -4m^2} \sim a^2 ,\\
 \langle \psi_{1,\rm int} \psi_{1,\rm int} \rangle
   \sim \langle \psi_{2, \rm int} \psi_{2, \rm int} \rangle
  &\sim \int^{2/a} d\hat{p} \frac{a \hat{p}}{\hat{p}^2-4m^2}\sim a, \\
 \langle \psi_{1, \rm int} \psi_{2, \rm int} \rangle
  &\sim \int^{2/a} d\hat{p} \frac{am}{\hat{p}^2-4m^2}  \sim a^2.
\end{align}
The diagonal propagators contribute with a factor $a$ in the
UV region, off-diagonal ones with a factor $a^2$. Considering now
an amputated diagram with $V$ vertices and $I$ internal lines of
which $I_{\rm off}$ have an off-diagonal propagator, its UV
contribution will be:
\begin{align}
 (\text{the total UV contribution})
  &\sim a^{I + I_{\rm off}} a^{V-1}a^{-V}
   \sim  a^{I+I_{\rm off}-1},
\end{align}
where the extra factor $a^{V-1}$ comes from the delta functions
which reduces the number of the momentum integrations. This
calculation shows that the superficial degree of divergence is
$I+I_{\rm off} -1$, which can give a logarithmic divergence only
for $I=1$, $I_{\rm off}=0$. However if we consider that a factor
$a^{-V}$ comes from the vertices, the contribution of the momentum
integration alone is given by:
\begin{align}
 (\text{the integration only}) &\sim a^{I + I_{\rm off}} a^{V-1},
 \label{momentumuv}
\end{align}
that is momentum integrations are always convergent in the UV.

How about the IR divergences?
All the propagators are convergent in the IR.  All the interactions are
finite as well.  Note that $\phi^n$ term in the $n$-point interaction
has a factor of $1/a$ but it is compensated by $(1-\cos\frac{ap}{2})$ in
the IR. Therefore momentum intergrations are always convergent in the IR
as well.

We are now in position to prove the main result,
namely that translational invariance is recovered in the continuum limit.
Since the conserved quantity is not the momentum itself $p$ but $\sin\frac{ap}{2}$
finite translational invariance is explicitly broken at a finite
lattice spacing. Indeed, if we denote the component fields by $\phi_A =
(\varphi, D, \psi_1, \psi_2)$, the sine conservation law implies
that correlation functions are invariant under the transformation:
\begin{align}
 \phi_A(p) &\rightarrow \exp(i l \frac{2}{a}\sin\frac{ap}{2}) \phi_A(p)
 \qquad l:\mbox{a finite length}
 \label{eq:finite-translation-sin}
\end{align}
whereas invariance under finite translation would require the invariance under the transformation
\begin{align}
 \phi_A(p) &\rightarrow \exp(i lp)\phi_A(p).\label{fint}
\end{align}
To prove that invariance under finite translations is recovered we need to
prove that in the continuum limit (\ref{eq:finite-translation-sin}) and (\ref{fint}) are equivalent.
For an $n$-point correlation function of $\phi_A $,
transformation (\ref{eq:finite-translation-sin}) is equivalent to
\begin{align}
 \langle \phi_{A_1}(p_1) \phi_{A_2}(p_2) \dots \phi_{A_n}(p_n) \rangle
 &\rightarrow
  \exp\left(\sum_{i=1}^n \frac{2l}{a} \sin\frac{ap}{2}\right)
 \langle \phi_{A_1}(p_1) \phi_{A_2}(p_2) \dots \phi_{A_n}(p_n) \rangle \nonumber \\
 &\simeq
 \left( 1 -i \frac{a^2l}{24}\sum_{i=1}^{n}p_i^3\right)
 \exp\left(il\sum_{i=1}^n p_i\right)
 \langle \phi_{A_1}(p_1) \phi_{A_2}(p_2) \dots \phi_{A_n}(p_n) \rangle. \label{compr}
\end{align}
where in the last step higher order terms in the expansion of $\sin\frac{ap}{2}$ have been neglected since $ap \rightarrow 0$ in the continuum limit.
The leading term that breaks translational invariance is then given by the second term in the bracket at the r.h.s. of (\ref{compr}). This vanishes as $a^2$ in the continuum limit if we assume $l p_i$ to be of order $1$ so that this term can be neglected as long as no divergence ( of order at least $\frac{1}{a^2}$ ) arises in the correlation  function $\langle \phi_{A_1}(p_1) \phi_{A_2}(p_2) \dots \phi_{A_n}(p_n) \rangle$.
As shown in the first part of this section this is not the case, so we can conclude that invariance under finite translations is recovered in the continuum limit.

\subsection{Ward-Takahashi identities}

Invariance under supersymmetry transformations is exact at the
finite lattice spacing and it is not spoiled by radiative
corrections, which are all finite in the lattice theory. Since the
continuum limit is smooth, we expect that  exact supersymmetry is
preserved also in this limit. This can be confirmed explicitly, by
checking that the corresponding Ward-Takahashi identities (WTi)
are satisfied. We shall consider here the case of a four points
interaction and check that the  supersymmetric Ward-Takahashi
identities are satisfied at 1-loop level. For the 2-point
correlation function, there are 3 independent WTis:
\begin{align}
 \cos\frac{ap}{4} \langle \psi_1(p) \psi_1(-p) \rangle
  + \frac{4}{a}\sin\frac{ap}{4} \langle \varphi(-p) \varphi(p)\rangle
 &=0, \label{eq:wti1}\\
 \frac{4}{a}\sin\frac{ap}{4} \langle \psi_1(p) \psi_1(-p) \rangle
  + \cos\frac{ap}{4} \langle D(-p) D(p)\rangle
 &=0,\label{eq:wti2}\\
 i \langle \psi_1(p) \psi_2(-p) + \langle \varphi(p) D(-p) \rangle
 &=0. \label{eq:wti3}
\end{align}
There are also  2 more identities obtained from the above by
replacing $\psi_1\to\psi_2,\ \psi_2\to-\psi_1$, but they are
automatically satisfied if  (\ref{eq:wti1}) and (\ref{eq:wti2})
are satisfied.

%Since the momentum is conserved in terms of the $\sin\frac{ap}{2}$
%instead of $p$, it is convenient to introduce
%\begin{equation}
% \hat{p} \equiv \frac{2}{a}\sin\frac{ap}{2}.
%\end{equation}
%This change of the variable gives an extra factor
%$\cos\frac{ap}{2}$ in the propagators, which becomes
%\begin{align}
% \langle \varphi'(\hp) \varphi'(-\hp) \rangle
%  &=\frac{c(\hp)}{\hp^2 -4m^2}\frac{1}{2}\left(1+c(\hp)\right),
%  \label{eq:propagatorp-phiphi}\\
% \langle D'(\hp) D'(-\hp) \rangle
%  &=\frac{c(\hp)}{\hp^2 -4m^2}\frac{8}{a^2}\left(1-c(\hp)\right),
%    \label{eq:propagatorp-DD}\\
% \langle \varphi'(\hp) D'(-\hp) \rangle
%  &=\frac{c(\hp)}{\hp^2 -4m^2} 2m,     \label{eq:propagatorp-phiD}\\
%  \langle \psi_1'(\hp)\psi_1'(-\hp) \rangle
%  =\langle \psi_2'(\hp)\psi_2'(-\hp) \rangle
%  &= - \frac{c(\hp)}{\hp^2-4m^2} \hp, \label{eq:propagatorp-psi1psi1}\\
% \langle \psi_1'(\hp)\psi_2'(-\hp) \rangle
%  = - \langle \psi_2'(\hp)\psi_1'(-\hp) \rangle
%  &= \frac{c(\hp)}{\hp^2-4m^2} 2im. \label{eq:propagatorp-psi1psi2}
%\end{align}
%Here, the primed fields are defined as
%\begin{align}
% \varphi'(\hp) &\equiv \varphi(p(\hp))
%  = \varphi\left(\frac{2}{a}\arcsin\frac{a\hp}{2}\right)\quad \text{etc.,}
%\end{align}
%and
%\begin{align}
% c(\hp)&\equiv \sqrt{1-\frac{a^2\hp^2}{4}} =\cos\frac{ap}{2}.
%\end{align}
At the tree level, it is easy to see that the WTi
(\ref{eq:wti1})--(\ref{eq:wti3})
are satified using propagators
(\ref{eq:propagatorp-phiphi})--(\ref{eq:propagatorp-psi1psi2}).

The 1-loop radiative corrections to these propagators are
calculated in  Appendix \ref{app:1-loop}. The result is:
\begin{align}
 \langle \varphi'(\hp)\varphi'(-\hp)\rangle_\text{1-loop}
  &= \langle \varphi'(\hp)\varphi'(-\hp)\rangle_\text{tree} F_1(\hp),\\
 \langle D'(\hp)D'(-\hp)\rangle_\text{1-loop}
  &= \langle D'(\hp)D'(-\hp)\rangle_\text{tree} F_1(\hp),\\
 \langle D'(\hp)\varphi'(-\hp)\rangle_\text{1-loop}
 &= \langle D'(\hp)\varphi'(-\hp)\rangle_\text{tree} F_2(\hp),
\end{align}
where
\begin{align}
 F_1(\hp)&=-ig8a^3 \frac{1}{D(\hp)}
      \int_{-2/a}^{2/a}\frac{dk}{2\pi}\frac{1}{D(k)}
       2(1+am) \left[(1+am)^2 -c^2(\hp) \right], \\
 F_2(\hp)&= -ig8a^3 \frac{1}{D(\hp)}
      \int_{-2/a}^{2/a}\frac{dk}{2\pi}\frac{1}{D(k)}
       \frac{2(1+am)}{am} \left[(1+am)^2 - (1+2am)c^2(\hp) \right]
\end{align}
with\footnote{Do not confuse with the auxiliary field.}
\begin{equation}
 \frac{1}{D(\hp)}=\frac{c(\hp)}{(a^2\hp^2-4a^2m^2)}.
\end{equation}
So 1-loop radiative corrections to the diagonal (resp.
off-diagonal) propagators are given by the function  $F_1(\hp)$
(resp.$F_2(\hp)$). The same thing happen in the case of fermionic
propagators:
\begin{align}
 \langle \psi_1'(\hp)\psi_1'(-\hp)\rangle_\text{1-loop}
  &= \langle \psi_1'(\hp)\psi_1'(-\hp)\rangle_\text{tree} F_1(\hp), \\
 \langle \psi'_1(\hp)\psi'_2(-\hp)\rangle_\text{1-loop}
  &= \langle \psi'_1(\hp)\psi'_2(-\hp)\rangle_\text{tree} F_2(\hp).
\end{align}
It follows that since the WT identities are satisfied at the tree
level they are also satisfied at 1-loop level.

\section{Leibniz rule and new star product in coordinate space \\
and the link approach}

Since we have established a new exactly supersymmetric lattice model, it
is instructive to compare the algebra of this model with that of link approach.
We first find out the algebraic structure of the model.
The momentum representation of supersymmetry transformations
(\ref{delta1pb},\ref{delta1pf}) and (\ref{delta2pb},\ref{delta2pf}) can be related to
the supercharges $Q_1$ and $Q_2$ as
\begin{align}
\delta_1 &= \alpha\sqrt{a} Q_1,   \nonumber \\
\delta_2 &= \alpha\sqrt{a} Q_2,  \label{delta_2-alg}
\end{align}
where the lattice constant dependence is introduced to recover the dimensionality
of supercharges. We find the following algebraic relation:
\begin{align}
Q_1^2 = Q_2^2 =  \frac{2}{a} \sin \frac{ap}{2},~~~~~~\{Q_1,Q_2\} =0. \label{delta-alg-mom}
\end{align}
The coordinate representation of the super charges $Q_1$ and $Q_2$ for the
supersymmetry transformations (\ref{delta1b} ,\ref{delta1f}) and
(\ref{delta2b} ,\ref{delta2f}) can be defined exactly same as
(\ref{delta_2-alg}), then we find the following supersymmetry algebra:
\begin{align}
Q_1^2 = Q_2^2 = i \hat{\partial}, ~~~~~~\{Q_1,Q_2\} =0, \label{delta-alg-co}
\end{align}
where the symmetric difference operator $\hat{\partial}$ is defined as:
\begin{align}
\hat{\partial} F(x) \equiv  \frac{(\partial_+ -\partial_-)}{a} F(x)
=\frac{F(x+\frac{a}{2}) -F(x-\frac{a}{2})}{a},     \label{diff-op}
\end{align}
with $\partial_{\pm}$ given in (\ref{fdiff}).
We find the following algebraic correspondence between the momentum representation
of derivative operator and the coordinate counterpart of difference operator:
\begin{align}
\frac{2}{a} \sin \frac{ap}{2}  \longleftrightarrow  i \hat{\partial}.     \label{momentum-op}
\end{align}
This lattice version of supersymmetry algebra coincides with the
continuum algebra of (\ref{cont-susy-alg1}) in the continuum limit $ap \rightarrow 0$.
As we have stressed in section 2, the lattice counter part of momentum operator
generates the lattice constant $a$ step translation of fields although the
basic lattice structure of this model has half lattice nature.

We have pointed out that the supersymmetry transformation is essentially
the half lattice spacing translation of lattice superfields with an
alternating sign factor as we can see in (\ref{st}).
The operation of the lattice half
step translation needs special care since
\begin{align}
\partial_{\pm} (-1)^{\frac{2x}{a}} \ne  (-1)^{\frac{2x}{a}}\partial_{\pm}.
\end{align}
%itself may not be the generator of the lattice translation.
%Therefore the forward and backward difference operators $\partial_{\pm}$
%defined in (\ref{fdiff}) do not neccessarily commute with super charges.
%In fact we find
%\begin{align}
%\delta_1 \partial_{\pm} = \partial_{\pm} \delta_1, ~~~~~~~
%\delta_2 \partial_{\pm} \ne \partial_{\pm} \delta_2.
%\end{align}
On the other hand the translation generator $\hat{\partial}$ commutes with
supersymmetry generators:
\begin{align}
\delta_1 \hat{\partial} = \hat{\partial} \delta_1, ~~~~~~~
\delta_2 \hat{\partial} = \hat{\partial} \delta_2,
\end{align}
which are equivalent to
\begin{align}
[Q_1, \hat{\partial}] = [Q_2 \hat{\partial}] = 0.
\end{align}
This leads to the continuum algebra (\ref{cont-susy-alg2}) in the continuum
limit. The full lattice constant spacing differential operator is the translation
generator, which is consistent with (\ref{delta-alg-co}).
We have thus confirmed that the lattice supersymmetry algebra of this
model leads to the continuum algebra in the continuum limit.
We can, however, recognize that lattice supersymmetry algebra has the same
form with the continuum super algebra even with a finite lattice constant.

We have shown already that the lattice version of this model have exact
supersymmetry at least in the momentum representation even with the
interaction terms. The coordinate formulation of the model should have
exact supersymmetry as well since it should in principle be equivalent with
the formulation of momentum representation. On the other hand the lattice
supersymmetry algebra of this model includes symmetric difference operator
as in (\ref{delta-alg-co}).
It is a well known fact that the symmetric difference operator (\ref{diff-op})
does not satisfy Leibniz rule for the product of fields:
\begin{align}
\hat{\partial}(F(x)G(x)) &= \frac{1}{a}\left(F(x+\frac{a}{2})
G(x+\frac{a}{2}) -  F(x-\frac{a}{2}) G(x-\frac{a}{2})\right) \nonumber \\
&= \hat{\partial}F(x) G(x+\frac{a}{2})
+ F(x-\frac{a}{2})\hat{\partial}G(x) \nonumber \\
                         &= \hat{\partial}F(x) G(x-\frac{a}{2})
+ F(x+\frac{a}{2})\hat{\partial}G(x). \label{symm-diff-Leip}
\end{align}
Here comes a crucial question: \\
{\it
``How can the lattice supersymmetry algebra be consistent
since the difference operator does not satisfy Leibniz rule while
the super charges satisfy Leibniz rule ?''
}\\

In the link approach of the lattice supersymmetry formulation this problem
was avoided by introducing shifting nature for the super charges:
\begin{align}
Q_1(F(x)G(x)) &= Q_1F(x) G(x+\frac{a}{4})
+ F(x-\frac{a}{4})Q_1G(x), \nonumber \\
Q_2(F(x)G(x)) &= Q_2F(x) G(x-\frac{a}{4})
+ F(x+\frac{a}{4})Q_2G(x) \label{Q-link},
\end{align}
where $F$ and $G$ are assumed to be bosonic lattice superfields. In the case
of fermionic superfields it works same if the anticommuting Grassmann
nature is taken into account.
We can confirm that the lattice supersymmetry algebra (\ref{delta-alg-co}) is
consistently fulfilled.
There is, however, ordering ambiguity for the product of fields:
\begin{align}
Q_1(F(x)G(x)) &= Q_1F(x) G(x+\frac{a}{4})
+ F(x-\frac{a}{4})Q_1G(x) \nonumber \\
= Q_1(G(x)F(x)) &= Q_1G(x) F(x+\frac{a}{4})
+ G(x-\frac{a}{4})Q_1F(x) \label{link-incons},
\end{align}
since $F(x)G(x)=G(x)F(x)$. We obtain a similar relation for $Q_2$.
Since the right hand sides of (\ref{link-incons}) are different, this
discrepancy was criticized as an inconsistency of the
formulation~\cite{{Dutch},{B-C-K}}.
It was, however, recognized that if there is a mild noncommutativity between
fields having a shifting nature:
\begin{align}
Q_1F(x) G(x+\frac{a}{4}) &= G(x-\frac{a}{4})Q_1F(x), \nonumber \\
Q_1G(x) F(x+\frac{a}{4}) &= F(x-\frac{a}{4})Q_1G(x),
\label{noncommut-rel}
\end{align}
where $F$ and $G$ are shiftless while $Q_1F$ and $Q_1G$ carry a shift of
$-\frac{a}{2}$, then there is no inconsistency. This algebraic consistency
was carefully investigated and it was discovered that these algebraic
relations (\ref{symm-diff-Leip}),(\ref{Q-link}) and (\ref{noncommut-rel})
are consistently treated by Hopf algebraic symmetry~\cite{DKS}.
Thus we may claim that the model has exact Hopf algebraic lattice
supersymmetry.

%In the current model, however, we have not introduced the shifting nature for
%the super charges and fields. The posed question still remains.
%In section 4 this question was partially answered. Here we make a careful
%look on the problem.

The exact invariance of the momentum representation of the
action (\ref{kterm}) under the supersymmetry transformations
(\ref{delta1pb}, \ref{delta1pf}) and (\ref{delta2pb}, \ref{delta2pf})
suggests that there should be coordinate counterpart which reflect
this exact invariance including interactions.
%For the kinetic terms and the mass terms supersymmetry invariance
%can be explicitly shown in the coordinate representation.
In the proof of the supersymmetry invariance, Leibniz rule is
used for the operation of super charges $Q_j$. It then leads to the
following relation:
\begin{align}
Q^2_j(F(x) * G(x)) &= (Q^2_jF(x)) * G(x)
+ F(x) * (Q^2_jG(x)),
\label{Q2-*Leibniz}
\end{align}
or equivalently
\begin{align}
\hat{\partial}(F(x) * G(x)) &= (\hat{\partial}F(x)) * G(x)
+ F(x) * (\hat{\partial}G(x)),
\label{deriv-*Leibniz}
\end{align}
where we have tentatively introduced a new type of $*$ product which satisfies
Leibniz rule even for the difference operator $Q^2_j=\hat{\partial}$ which
is Euclidean version of (\ref{delta-alg-co}) and
normally satisfies shifted Leibniz rule (\ref{symm-diff-Leip}) for the normal
product of fields.
%If we can find out a consistent solution of the $*$ product
%the puzzle is solved.
However in the proof of the supersymmetry invariance of the kinetic terms
and the mass terms in the coordinate representation Leibniz rule has been
used for the normal products and thus the relations (\ref{Q2-*Leibniz})
and (\ref{deriv-*Leibniz}) should hold for the normal products,
at least for the product of two fields, which seems
to be inconsistent with the shifted Leibniz rule of symmetric difference
operator (\ref{symm-diff-Leip}).
This is rephrasing the puzzle of the current problem.

Assuming that the Leibniz rule for the difference operator works for the
normal product, we find the following difficulty.
That is, supose we had defined $\hat{\partial}$ operation on a product
of fields as
\begin{equation}
 \hat{\partial}(F(x)G(x)H(x)\cdots )
 \overset{?}{=}
  (\hat{\partial}F(x))G(x)H(x)\cdots
 +F(x)(\hat{\partial}G(x))H(x)\cdots
 +F(x)G(x)(\hat{\partial}H(x))\cdots.
 \label{leibniz-questioned}
\end{equation}
This new definition does not necessarily lead to a cancellation of surface
terms for the product of superfields and thus supersymmetry cannot be kept
exactly.
Up to the product of two superfields, the surface terms cancel using the
r.h.s of (\ref{leibniz-questioned})
\begin{align}
\sum_{x}\hat{\partial}F(x) = 0, ~~~~~~\sum_{x} \hat{\partial}(F(x)G(x))
=0.
\label{surface-sum-1}
\end{align}
However, in general the surface terms for a product of more than three
superfields do not cancel:
\begin{align}
\sum_{x} \hat{\partial}(F(x)G(x)H(x)\cdots) \ne 0.
\label{surface-mult-prod}
\end{align}
We must find a formulation of a new $*$ product which satisfies the
$*$ product version of (\ref{surface-mult-prod}) where the nonequality
changes to equality:
\begin{align}
\sum_{x} \hat{\partial}(F(x)*G(x)*H(x)\cdots) = 0.
\label{surface-mult-star-prod}
\end{align}

We have recognized in the previous sections that non locality
plays an important role in the present formulation. We have also
recognized that non locality stems from the sine momentum
conservation (\ref{sincons}) which, in turn, arises from the
necessity on the lattice to have complete periodicity in the
momentum and to have the species doublers on the same footing as
the original fields. With ordinary momentum conservation the
product of a field $F$ of momentum $p_1$ and a field $G$
of momentum $p_2$ is a composite field $\Phi=F \cdot G$
of momentum $p=p_1+p_2$, namely the momentum is the additive
quantity under product: \beq \Phi(p) \equiv (F\cdot G)(p)
= \frac{a}{2\pi}\int d p_1 d p_2 F(p_1) G(p_2)
\delta(p-p_1-p_2). \label{dotprod} \eeq
 In coordinate space this amounts to the ordinary local product:
 \beq
 \Phi(x) \equiv (F \cdot G)(x) = F(x) G(x).
 \label{dotcoord}
 \eeq
 On the lattice momentum conservation is replaced by the
lattice (sine) momentum conservation (\ref{sincons}), which means that $\hat{p}= \frac{2}{a} \sin\frac{a p}{2}$ is the additive quantity when taking the product of two fields. In other words  the product of a field $F$ of momentum $p_1$ and a field $G$ of momentum $p_2$ is a composite field $\Phi=F * G$ of momentum $p$ with $\sin\frac{ap}{2}=\sin\frac{a p_1}{2}+\sin\frac{a p_2}{2}$.
 This amounts to changing the definition of the ``dot'' product  to that of a ``star'' product defined in momentum space as
 \footnote{With this definition the star product is periodic in $p$ with period $\frac{4 \pi}{a}$.
 So while it is suitable for bosons, in order to apply it to fermions we have to redefine the fermion fields
 and make them periodic. This can be done by defining $\Psi_p(p)= e^{-ipa/4}\Psi(p)$ and use $\Psi_p $ in the
 definition of the ``star'' product. $\Psi_p(p)$ satisfies the reality condition $\Psi_p^\dagger(p) =\Psi_p(-p)$
 and can be used in the action and in the SUSY transformations instead of $\Psi(p)$. The main difference is that with
 the use of $\Psi_p$ fermions are like the bosons at the sites $x=n\frac{a}{2}$ in the coordinate representation
 and the SUSY transformations are expressed in terms of right (or left) finite differences of spacing $\frac{a}{2}$
 instead of the symmetric one as in (\ref{delta1b},\ref{delta1f}).}  :
 \beq
 \Phi(p) \equiv (F * G)(p)
= \frac{a}{2\pi}\int d \hat{p}_1 d \hat{p}_2 F(p_1) G(p_2) \delta(\hat{p}-\hat{p}_1-\hat{p}_2)   \label{starprod}
\eeq As we shall see this product is not anymore local in
coordinate space but satisfies the Leibniz rule with respect to
the symmetric difference operator $\hat{\partial}$. This is easily
checked in the momentum representation. In fact, according to
(\ref{momentum-op}) acting with $\hat{\partial}$ corresponds in
momentum space to multiplication by
$\hat{p}=\frac{2}{a}\sin\frac{ap}{2}$, so that from
 (\ref{starprod}) we get: \beq \hat{p}~ \Phi(p)
 = \frac{a}{2\pi}\int ~d \hat{p}_1~ d \hat{p}_2 \left[ \hat{p}_1~ F(p_1) ~
G(p_2)+ F(p_1)~\hat{p}_2~G(p_2) \right]
\delta(\hat{p}-\hat{p}_1-\hat{p}_2).   \label{lbrule} \eeq

Explicit form of the coordinate representation of the star product
is given by
\begin{align}
(F*G) (x)
&= F(x) * G(x)
=a\int \frac{d\hat{p}}{2\pi}\, e^{-ipx}~ (F*G)(p) \nonumber \\
&=\int_{-\frac{\pi}{2}}^{\frac{3\pi}{2}} d\tilde{p}~ \cos\tilde{p}
~e^{-ipx} \int_{-\frac{\pi}{2}}^{\frac{3\pi}{2}}
\frac{d\tilde{p}_1}{2\pi} \frac{d\tilde{p}_2}{2\pi}
\cos\tilde{p}_1 ~\cos\tilde{p}_2
\int_{-\infty}^\infty  \frac{d\tau}{2\pi}
e^{i{\tau}(\sin\tilde{p}-\sin\tilde{p}_1-\sin\tilde{p}_2)}
\nonumber \\
&~~~~~~~\times \sum_{y,z}e^{i(m\tilde{p}_1 +l\tilde{p}_2)} F(y) G(z)
\nonumber \\
&= \int_{-\infty}^\infty d\tau J_{n\pm1}({\tau})
\sum_{m,l}J_{m\pm1}({\tau})J_{l\pm1}(\tau) F(y) G(z),
\label{def-star-prod-cor}
\end{align}
where $\tilde{p}=\frac{ap}{2}$, %$\tilde{\tau}=\frac{2\tau}{a}$
and $x=\frac{na}{2}, y=\frac{ma}{2}, z=\frac{la}{2}$ should be understood and where the  integration variable is not $p$ but $\hat{p}$.

The lattice delta function is parametrized by $\tau$
\begin{align}
\delta\left(\frac{2}{a}\sin\tilde{p}_i\right)
=\frac{a}{4\pi}\int^\infty_{-\infty}
d\tau e^{i{\tau} \sin\tilde{p}_i}.
\label{deltafunction}
\end{align}
$J_n(\tau)$ is a Bessel function defined as
\begin{align}
J_n(\tau)=\frac{1}{2\pi}\int_\alpha^{2\pi+\alpha}
e^{i(n\theta-\tau \sin \theta)} d\theta,
\label{Bessel}
\end{align}
and we use the following notation:
\begin{align}
J_{n\pm1}(\tau)=\frac{1}{2}(J_{n+1}(\tau) + J_{n-1}(\tau)).
\label{def-Jpm}
\end{align}
It is obvious that the star product is commutative:
\beq
F(x) * G(x) = G(x) * F(x).
\label{star-commut}
\eeq
We can now check how the difference operator acts on the star product
of two lattice superfields and find that the difference operator action
on a star product indeed satisfies Leibniz rule:
\begin{align}
 i\hat{\partial} (F(x)*G(x)) &=
a\int \frac{d\hat{p}}{2\pi}~ i\hat{\partial}_x~e^{-ipx}~(F*G)(p)
\nonumber \\
%&=\int d\hat{p}~e^{-ipx}~\hat{p}~(F*G)(p) \nonumber \\
%&=\int d\hat{p}~e^{-ipx} \int d\hat{p}_1 d\hat{p}_2
%\left((\hat{p}_1 F(p_1))G(p_2) +
%F(p_1)(\hat{p}_2 G(p_2))\right)
%~\delta (\hat{p}-\hat{p}_1-\hat{p}_2)
%\nonumber \\
%&=\int d\hat{p}~e^{-ipx}\sum_{y,z}\int d\hat{p}_1d \hat{p}_2
%\nonumber \\
%&~~~\left((-i\hat{\partial}_y~e^{ip_1y})~e^{ip_2z} +
%e^{ip_1y}(-i\hat{\partial}_z~e^{ip_2z})\right)
%~F(y)~G(z)
%\delta (\hat{p}-\hat{p}_1-\hat{p}_2)
%\nonumber \\
&=\frac{a^2}{4} \int d\hat{p}~e^{-ipx}\sum_{y,z}\int
\frac{d\hat{p}_1}{2\pi} \frac{d\hat{p}_2}{2\pi}~e^{ip_1y+ip_2z}
\nonumber \\
&~~~~~~~\times \left((i\hat{\partial}_y~F(y))G(z)~+~
F(y)~(i\hat{\partial}_z~G(z)) \right)
~\delta (\hat{p}-\hat{p}_1-\hat{p}_2)\nonumber \\
&=(i\hat{\partial} F(x))*G(x) + F(x) * (i\hat{\partial} G(x)).
\label{star-Leibniz}
\end{align}
Eq. (\ref{star-Leibniz}) already answer the question posed at the
beginning of this section: the Leibniz rule for the symmetric
finite difference operator $\hat{\partial}$ is recovered by the
redefinition of the product of fields, in agreement with the sine
momentum conservation on the lattice.

Similar to the case for star product of two fields we can
show that the difference operator acting on the star product of three
fields satisfies the Leibniz rule and the surface terms of a star product
for more than 3 lattice superfields vanishes:
\begin{align}
\sum_x i\hat{\partial}( F(x)*G(x)*H(x))
&=\sum_x \left( (i\hat{\partial}F(x) )*G(x)*H(x)+
F(x)*(i\hat{\partial}G(x))*H(x) \right. \nonumber \\
&~~~~+\left. F(x)*G(x)*(i\hat{\partial}H(x) ) \right)
\nonumber \\
&= \sum_x i\hat{\partial} \int d\hat{p}~ e^{-ipx}~(F*G*H)(p) \nonumber \\
&=\int dp \cos\tilde{p}~\delta(p)~ \sin\tilde{p}~(F*G*H)(p) \nonumber \\
&= 0,
\label{surface-3prod}
\end{align}
where we have used the following relation:
\begin{align}
\sum_x e^{-ipx} = \frac{4\pi}{a}\delta (p).
\label{delta-func}
\end{align}
We are considering that our lattice coordinate space has
infinite extension and thus the lattice is discrete but the momentum
can be continuous.
The relation (\ref{surface-3prod}) works exactly similar to a star product
of more than 3 fields.
Thus this star product has the desired property of
(\ref{surface-mult-star-prod}).

After defining the new star product we may look at the kinetic and mass terms.
The $S^{(2)}$ without $g_0^{(2)}$ but with the regularization factor
of (\ref{cosfunc}) is given by
\begin{align}
 S^{(2)}
 &=2a^2 \int_{-\frac{\pi}{a}}^{\frac{3\pi}{a}} \frac{d p_1}{2\pi}
\frac{d p_2}{2\pi}
   2\pi\delta\left(\sum_{i=1}^2\sin\frac{a p_i}{2}\right)
\cos\frac{ap_1}{2} \cos\frac{ap_2}{2}
   \nonumber\\
 &\times \Bigl[
  (1-\cos\frac{a p_1}{2})~\Phi(p_1)\Phi(p_2) +
   \frac{1}{2}~ \sin\frac{a(p_1-p_2)}{4}~ \Psi(p_1)\Psi(p_2)\Bigr]
\nonumber \\
&= \sum_{x}\Bigl[\Phi(x)* \left(2\Phi(x)
- \Phi(x+\frac{a}{2})-\Phi(x-\frac{a}{2})\right)
+\frac{i}{2} \Psi(x+\frac{3a}{4}) * \Psi(x+\frac{a}{4})\Bigr].
%\nonumber \\
%&= S_{kin} +\frac{1}{m_0}S_{mass},
  \label{S2*term}
\end{align}
%where $ S_{kin}$ and $S_{mass}$ are the kinetic and mass terms of local
%actions given in (\ref{coordskin}) and (\ref{coordsmass}) respectively.
%Here the sine conservation is replaced by the sum of two delta functions
%with the replacement in (\ref{deltas}) with $m_0=1$.
It is interesting to recognize that the coordinate representation of
the action with star product has almost the same form of the kinetic
term of the local action, $S_{kin}$ in (\ref{coordskin}),
where the star product is just replaced by the
normal product.
The arguments of the fermionic lattice superfield in
$S_{kin}$ is shifted with $\frac{a}{2}$ from that of (\ref{S2*term}).
This is due to the loss of lattice translational
invariance in the star product formulation while in the local expression
the lattice translational invariance is recovered and thus a half lattice
shift is equivalent in the action.
This equivalent form correspondence between the star product action and
the kinetic term of the local action leads to a recognizability that
the star product action is invariant under the supersymmetry
transformations by $Q_1$ and $Q_2$ acting on the star product
of fields by keeping Leibniz rule since it is invariant for
the local action.
This correspondence in return to a result that the difference
operator acting on local fields should satisfies Leibniz rule since
the difference operator acting on the star product fields satisfies
Leibniz rule. This solves the puzzle of the problem posed in this section.

This $S^{(2)}$ action in the star products form, however, is equivalent
to a sum of both the kinetic terms and the mass terms with fixed
coefficient, which include product of local fields.
In deriving the local action of the $S_{kin}$ in (\ref{coordskin})
and $S_{mass}$ in (\ref{coordsmass}) the regularization factor for
lattice momentum conservation (\ref{cosfunc}) has not been included.
In the star product formulation of lattice field theory the
regularization factor is automatically involved since the lattice
momentum itself $\hat{p}=\frac{a}{2}\sin \frac{ap}{2}$ is the integration
variable.

Similarly to $S^{(2)}$ we can now derive the coordinate representation of the
general interaction action $S^{(n)}$.
We first note the following relation:
\begin{align}
2\pi\delta\left(\sum_{j=1}^n\sin\frac{a p_j}{2}\right)
&= \frac{a}{2}\sum_x\int d(\sin\tilde{p}) e^{-ipx}\delta\left( \sin\tilde{p}
-\sum_{j=1}^n\sin\tilde{p}_j \right).
\label{delta-relation}
\end{align}
Then the general interaction action (\ref{nterm}) can be given by the
following form:
\begin{align}
 S^{(n)} &=\frac{4g_0^{(n)}}{n!(2\pi)^n} \sum_x \int d\tilde{p} \cos\tilde{p}
e^{-ipx} \int \prod_{j=1}^n (d\tilde{p}_j\cos\tilde{p}_j) \int d\tau
e^{i\tau(\sin\tilde{p}-\sum_{j=1}^n\sin\tilde{p}_j )}
\!\sum_{y_1,y_2, \cdots y_n}\! e^{i(\sum_{j=1}^n \tilde{p}_j m_j)}
\nonumber \\
& \Bigl[
  \left(1-\frac{1}{2}(e^{i\tilde{p}_1} + e^{-i\tilde{p}_1} )\right)
\Phi(y_1)\Phi(y_2)\cdots\Phi(y_n)
\nonumber \\
&~~~~~~~~~~~~~~~~~~~~~~~~~~~~~~~-\frac{(n-1)i}{8}(e^{i\tilde{p}_1}
-e^{i\tilde{p}_2})
\Psi(y_1+\frac{a}{4})\Psi(y_2+\frac{a}{4})\Phi(y_3)\cdots \Phi(y_n)\Bigr]
\nonumber \\
%&= 2^ng_0^{(n)}\frac{2\pi}{a}\sum_x\int^\infty_{-\infty} d\tau
%J_{n\pm1}(\tau)\sum_{m_1,\cdots,m_n}\prod_{j=1}^n J_{m_j\pm1}(\tau)
%\nonumber \\
%& \Bigl[\left(2\Phi(y_1)
%- \Phi(y_1+\frac{a}{2})-\Phi(y_1-\frac{a}{2})\right)
%\Phi(y_2)\cdots \Phi(y_n)
%-\frac{(n-1)i}{2} \Psi(y_1+\frac{3a}{4}) \Psi(y_2+\frac{a}{4}) \Phi(y_3)
%\cdots \Phi(y_n) \Bigr]
%\nonumber \\
&=\frac{4}{n!}g_0^{(n)} \sum_x
\Bigl[\left(2\Phi(x)
- \Phi(x+\frac{a}{2})
-\Phi(x-\frac{a}{2})\right)* \Phi(x)^{n-1}(x)
\nonumber \\
&~~~~~~~~~~~~~~~~~~~~~~~~~~~~~~~+\frac{(n-1)i}{2} \Psi(x+\frac{3a}{4})*
\Psi(x+\frac{a}{4})* \Phi^{n-2}(x) \Bigr],
\label{star-Sn-action}
\end{align}
where the relations: $x=\frac{na}{2}, \tilde{p_j} = \frac{p_ja}{2},
y_j=\frac{m_ja}{2}$
should be understood and $J_{n\pm1}(\tau)$ is given in (\ref{def-Jpm}).
$\Phi^{n}(x)$ is a star product of $n$ bosonic superfields.
We have defined the star product of $n$ fields as:
\begin{align}
F_1(x+b_1)* F_2(x+b_2) * \cdots *F_n(x+b_n) = %\frac{2}{a}
 \int^\infty_{-\infty} d\tau
J_{n\pm1}(\tau)\!\sum_{m_1,\cdots,m_n}\!
\left(\prod_{j=1}^n J_{m_j\pm1}(\tau)F_j(y_j+b_j)\right).
\label{star-n-prod}
\end{align}

The non local nature of the star product should disappear in the
continuum limit. This is however non trivial due to the $p
\rightarrow \frac{2 \pi}{a} - p$ symmetry of the $\sin\frac{a
p}{2}$ function and the existence of two translationally invariant
vacua at $p=0$ and $p=\frac{2\pi}{a}$. It was shown by Dondi and
Nicolai~\cite{D-N} that in the continuum limit namely at fixed $x$
with $a\rightarrow 0$: \beq J_{\frac{2 x}{a}} (\tau) \rightarrow
\delta(\tau-\frac{2 x}{a}).
  \label{contlimit}
\eeq However in the present context the continuum limit picks up
also the configuration at $p= \frac{2\pi}{a}$ and the previous
result has to be replaced by: \beq J_{\frac{2 x}{a}} (\tau)
\rightarrow \delta(\tau-\frac{2 x}{a})+ (-1)^{\frac{2
x}{a}}\delta(\tau+\frac{2 x}{a}). \label{contlim2} \eeq Thus
locality is recovered in the continuum limit, but with an extra
coupling of fields in the points $x$ and $-x$ accompanied with the
alternating sign factor $(-1)^{\frac{2x}{a}}$. Such remaining non
locality disappears when the lattice field $\Phi$ and $\Psi$ are
reinterpreted in terms of component fields as discussed in the
previous section.

We have now found a consistent definition of supersymmetry algebra
in the coordinate space as well by the star product which
assures the Leibniz rule operation of difference operator.
The supercharge operation for star product of fields
satisfies Leibniz rule and has the following form:
\begin{align}
Q_j(F(x)*G(x)) &= Q_jF(x) * G(x)
+ (-1)^{|F|}F(x) * Q_jG(x) \label{Q-star-product}.
\end{align}
Thus the operation of supersymmetry charges and translation
generators on lattice superfields are consistently
defined as an algebra both in momentum and coordinate representation.

We finally comment on an interesting possibility: \\
{\it ``The formulation of the present model with lattice momentum
conservation, equivalently the star product formulation of
lattice theory, and that of link approach are equivalent.'' }\\
This is based on the following observation: the algebraic correspondence of
$\hat{\partial}$ and $Q_j$, respectively,
(\ref{deriv-*Leibniz}) and
(\ref{Q-star-product}) with respect to (\ref{symm-diff-Leip}) and
(\ref{Q-link})
are exactly parallel and algebraically both of frameworks
have one to one correspondence
if the mild noncommutativity of (\ref{noncommut-rel}) is introduced in the
link approach. The current formulation of algebra is Lie algebraic
lattice supersymmetry with a new star product in the coordinate representation
while the link approach is Hopf algebraic lattice supersymmetry.
The nonlocality in the star product and the noncommutativity in the link
approach are corresponding.

\section{Conclusion and discussions}

We have proposed a new lattice supersymmetry formulation which ensures an
exact Lie algebraic supersymmetry invariance on the lattice for all super charges
even with interactions.
We have introduced bosonic and fermionic lattice superfields which accommodate
species doublers as bosonic and fermionic particle fields of super multiplets.
This lattice superfield formulation is, however, not a naive extension of continuum
superfield formulation in the sense that there appear higher order irrelevant terms,
including species doubler particle fields, which do not appear in the continuum
formulation because of dimensional reason.
These irrelevant terms are, however, crucial to assure the exact lattice
supersymmetry invariance. We consider that this lattice superfield formulation
is fundamental for a regularization of supersymmetry on the lattice.

As the simplest model we have explicitly investigated $N=2$ model
in one dimension. The model includes interaction terms and the
exact lattice supersymmetry invariance of the action for two
supersymmetry charges are shown explicitly. In the momentum
representation of the formulation the standard momentum
conservation is replaced by the lattice counterpart of momentum
conservation: the sine momentum conservation. The basic lattice
structure of this one dimensional model is half lattice spacing
structure and the lattice supersymmetry transformation is
essentially a half lattice spacing translation. The super
coordinate structure and the momentum representation of species
doubler fields is hidden implicitly in the alternating sign
structure of a half lattice spacing in the coordinate space. This
sign change with a half lattice spacing shift is a typical
phenomenon of lattice regularization and crucially related to the
both lattice supersymmetry and the chiral fermion regularization.
The sign change is a typical of lattice regularization and can
never appear in the continuum regularization and thus is
considered to be fundamental for the lattice supersymmetry.

Since we introduce the lattice (sine) momentum conservation: the translational
invariance is broken. We have investigated this problem and shown explicitly how
the translational invariance recovers in the continuum limit.
The Ward-Takahashi identity is investigated for a model with $\Phi^4$ interaction
term and it is confirmed that the identity is satisfied in one
loop level and is satisfied in all orders since this model is shown to be
super renormalizable.
In the previous investigation a fermionic species doubler
contribution from Wilson term, having inverse lattice constant mass,
was responsible to break the identity and the bosonic counter term was
necessary  to cancel this unwanted term~\cite{Giedt,CKU}.
In our model the species doubler contribution of fermion is identified
as a physical contribution as a super multiplet and this fermionic contribution can be compensated by the bosonic counterpart of species doubler.

Since the symmetric difference operator does not satisfy Leibniz rule,
it was very natural to ask how the supersymmetry algebra be consistent
in the coordinate space since super charges satisfy Leibniz rule.
In the link approach this problem was avoided by introducing shift
nature for super charges. In the current formulation this puzzle
is beautifully solved by introducing a new star product of lattice
superfields: The difference operator satisfies Leibniz rule on
the star products of lattice super fields. Then Lie algebraic
exact supersymmetry is realized in this coordinate formulation of star
product lattice field theory. This formulation provides a new
well defined regularization scheme of fermions and bosons without
species doubling problem of fermions. One may say otherwise,
{\it the regularization of fermions on the lattice inevitably
leads to supersymmetric lattice fields theories.}

It is recognized that the Lie algebraic structure of lattice
supersymmetry and the algebra of link approach
are totally one to one corresponding if mild
noncommutativity is introduced to the link approach. This suggests
an interesting possibility that
{\it the Lie algebraic lattice supersymmetry
and the Hopf algebraic supersymmetry of link approach are equivalent.}
The nonlocality in the star product and the noncommutativity in the
link approach are corresponding.
The confirmation of this interesting possibility will be left for
future investigation.

Since we have established a new lattice supersymmetry formulation
which has exact supersymmetry on the lattice, it would be important
to extend the formulation into higher dimensions and to the models
with gauge fields.

\vspace{1cm}

{\bf{\Large Acknowledgments}}

This work was supported in part by Japanese Ministry of Education, Science, Sports and Culture
under the grant number 50169778 and also by Insituto Nazionale di Fisica Nucleare (INFN)
research funds. I. Kanamori is financially supported by Nishina memorial
foundation.

\begin{center}
{\bf{\Large Appendix}}
\end{center}

\appendix
\section{Calculation of the 1-loop contribution to 2-point functions}
\label{app:1-loop}

Let us consider the four points interaction term. To write it down
it is convenient to introduce the following combination:
\begin{align}
 \phi_{\pm}(\hp) &\equiv
  \frac{1}{a}\left(\varphi'(\hp) \pm \frac{a}{4}D'(\hp)\right).
\end{align}
The 4 point interaction terms become
\begin{align}
 S_{\rm B}^{(4)}
 &= \frac{16}{3} g a^3 \int_{-2/a}^{2/a}
 \frac{d\hp_1}{2\pi}\frac{d\hp_2}{2\pi}\frac{d\hp_3}{2\pi}\frac{d\hp_4}{2\pi}
 2\pi\delta(\hp_1+\hp_2+\hp_3+\hp_4) \nonumber\\
 & \qquad \times \left[
  \phi_{+}(\hp_1)\phi_{+}(\hp_2)\phi_{+}(\hp_3)\phi_{+}(\hp_4)
 - c(\hp_1) \phi_{-}(\hp_1)\phi_{+}(\hp_2)\phi_{+}(\hp_3)\phi_{+}(\hp_4)
 \right], \\
 S_{\rm F}^{(4)}
 &=  8 ga^2 \int_{-2/a}^{2/a}
 \frac{d\hp_1}{2\pi}\frac{d\hp_2}{2\pi}\frac{d\hp_3}{2\pi}\frac{d\hp_4}{2\pi}
 2\pi\delta(\hp_1+\hp_2+\hp_3+\hp_4) \nonumber\\
 &\quad \times \Bigl[
  i\cos\frac{a(p_1+p_2)}{4}\psi'_1(\hp_1)\psi'_2(\hp_2) \phi_{+}(\hp_3) \phi_{+}(\hp_4)  \\
 & \qquad
  +\frac{1}{2}\sin\frac{a(p_1-p_2)}{4}\psi'_1(\hp_1)\psi'_1(\hp_2)\phi_{+}(\hp_3)\phi_{+}(\hp_4)  \\
 & \qquad
  +\frac{1}{2}\sin\frac{a(p_1-p_2)}{4}\psi'_2(\hp_1)\psi'_2(\hp_2)\phi_{+}(\hp_3)\phi_{+}(\hp_4)
\Bigr].
\end{align}
For the $S_{\rm F}^{(4)}$, the momenta insides the trigonometric
function is $p$ instead of $\hp$.

The propagators are
\begin{align}
 \langle \phi_{+}(\hp)\phi_{+}(-\hp) \rangle &=\frac{1+am}{D(\hp)}\\
 \langle \phi_{-}(\hp)\phi_{-}(-\hp) \rangle &=\frac{1-am}{D(\hp)}\\
 \langle \phi_{+}(\hp)\phi_{-}(-\hp) \rangle &=\frac{c(\hp)}{D(\hp)}\\
 \langle \psi'_1(\hp)\psi'_1(-\hp)\rangle
 = \langle \psi'_2(\hp)\psi'_2(-\hp)\rangle &=-\frac{a^2 \hp}{D(\hp)} \\
 \langle \psi'_1(\hp)\psi'_2(-\hp)\rangle
 =  -\langle \psi'_2(\hp)\psi'_1(-\hp)\rangle &=\frac{a^2 2im}{D(\hp)},
\end{align}
where
\begin{equation}
 \frac{1}{D(\hp)}=\frac{c(\hp)}{a^2\hp^2-4a^2m^2}.
\end{equation}

\newcommand{\myloopbb}{
\begin{fmffile}{1loop_diagram_bb}
\unitlength=1mm \hspace{\unitlength} \raisebox{-6\unitlength}{
\begin{fmfgraph*}(28,12)
 \fmfleft{i}\fmfright{o}
 \fmfv{decor.shape=cross,decor.size=4thick}{i}
 \fmfv{decor.shape=cross,decor.size=4thick}{o}
 \fmf{dashes}{i,v1}
 \fmf{dashes}{o,v1}
\fmfblob{.4w}{v1}
\end{fmfgraph*}
}
\end{fmffile}
}

\newcommand{\myloopff}{
\begin{fmffile}{1loop_diagram_ff}
\unitlength=1mm \hspace{\unitlength} \raisebox{-6\unitlength}{
\begin{fmfgraph*}(28,12)
 \fmfleft{i}\fmfright{o}
 \fmfv{decor.shape=cross,decor.size=4thick}{i}
 \fmfv{decor.shape=cross,decor.size=4thick}{o}
 \fmf{fermion}{i,v1}
 \fmf{fermion}{v1,o}
\fmfblob{.4w}{v1}
\end{fmfgraph*}
}
\end{fmffile}
}

Since the propagators are not diagonal, we first calculate
diagrams without contracting with the external fields. They are
\begin{align}
 \phi_{+}(\hp) \myloopbb \phi_{+}(-\hp)
  &= \phi_{+}(\hp)\phi_{+}(-\hp)
      (-ig) 8 a^3 \int_{-2/a}^{2/a}\frac{dk}{2\pi}\frac{1}{D(k)} 4(1+am),\\
 \phi_{+}(\hp) \myloopbb \phi_{-}(-\hp)
  &= \phi_{+}(\hp)\phi_{-}(-\hp)
      (-ig) 8 a^3 \int_{-2/a}^{2/a}\frac{dk}{2\pi}\frac{1}{D(k)}
      \left[-2(1+am)c(\hp)\right] ,\\
  \phi_{-}(\hp) \myloopbb \phi_{-}(-\hp)
   &=0\\
 \psi'_{1}(\hp) \myloopff \psi'_1(-\hp)
  &= \psi'_1(\hp)\psi'_1(-\hp)
      (-ig) 8 a^2 \int_{-2/a}^{2/a}\frac{dk}{2\pi}\frac{1}{D(k)}
     \frac{a\hat{p}}{2}(1+am), \\
 \psi'_2(\hp) \myloopff \psi'_2(-\hp)
  &= \psi'_2(\hp)\psi'_2(-\hp)
      (-ig) 8 a^2 \int_{-2/a}^{2/a}\frac{dk}{2\pi}\frac{1}{D(k)}
     \frac{a\hp}{2}(1+am), \\
 \psi'_1(\hp) \myloopff \psi'_2(-\hp)
  &= \psi'_1(\hp)\psi'_2(-\hp)
      (-ig) 8 a^2 \int_{-2/a}^{2/a}\frac{dk}{2\pi}\frac{1}{D(k)}
     i(1+am).
 \end{align}

After contracting with the external lines, we obtain the 2-point
functions for $\phi_{\pm}$ as
\begin{align}
 \langle \phi_{+}(\hp)\phi_{+}(-\hp)\rangle_\text{1-loop}
  &= -ig 8 a^3 \frac{1}{D(\hp)^2}\int_{-2/a}^{2/a}\frac{dk}{2\pi}\frac{1}{D(k)}
     4(1+am)\left[ (1+am)^2 - (1+am)c^2(\hp) \right], \\
 \langle \phi_{-}(\hp)\phi_{-}(-\hp)\rangle_\text{1-loop}
  &= -ig 8 a^3 \frac{1}{D(\hp)^2}\int_{-2/a}^{2/a}\frac{dk}{2\pi}\frac{1}{D(k)}
     4(1+am) am c^2(\hp), \\
 \langle \phi_{+}(\hp)\phi_{-}(-\hp)\rangle_\text{1-loop}
  &= -ig 8 a^3 \frac{1}{D(\hp)^2}\int_{-2/a}^{2/a}\frac{dk}{2\pi}\frac{1}{D(k)}
     2(1+am)c(\hp)\left[ (1+am)^2 - c^2(\hp) \right].
\end{align}
In terms of the scalar $\varphi$ and the auxiliary field $D$ these
become:
\begin{align}
 \langle \varphi'(\hp)\varphi'(-\hp)\rangle_\text{1-loop}
  &= \langle \varphi'(\hp)\varphi'(-\hp)\rangle_\text{tree} F_1(\hp),\\
 \langle D'(\hp)D'(-\hp)\rangle_\text{1-loop}
  &= \langle D'(\hp)D'(-\hp)\rangle_\text{tree} F_1(\hp),\\
 \langle D'(\hp)\varphi'(-\hp)\rangle_\text{1-loop}
 &= \langle D'(\hp)\varphi'(-\hp)\rangle_\text{tree} F_2(\hp),
\end{align}
where
\begin{align}
 F_1(\hp)&=-8iga^3 \frac{1}{D(\hp)}
      \int_{-2/a}^{2/a}\frac{dk}{2\pi}\frac{1}{D(k)}
       2(1+am) \left[(1+am)^2 -c^2(\hp) \right], \\
 F_2(\hp)&= -8iga^3 \frac{1}{D(\hp)}
      \int_{-2/a}^{2/a}\frac{dk}{2\pi}\frac{1}{D(k)}
       \frac{2(1+am)}{am} \left[(1+am)^2 - (1+2am)c^2(\hp) \right].
\end{align}
%The diagonal ones are accompanied by $F_1(q)$ and the off-diagonal one
%by $F_2(q)$.
For the fermionic 2-point functions, we obtain:
\begin{align}
 \langle \psi_1'(\hp)\psi_1'(-\hp)\rangle_\text{1-loop}
  &= \langle \psi_1'(\hp)\psi_1'(-\hp)\rangle_\text{tree} F_1(\hp), \\
 \langle \psi'_1(\hp)\psi'_2(-\hp)\rangle_\text{1-loop}
  &= \langle \psi'_1(\hp)\psi'_2(-\hp)\rangle_\text{tree} F_2(\hp).
\end{align}

%%%%%%%%%%%%%%%%%%%%%%%%%%%%%%%%%%%%%%%%%%%%%%%%%%%%%%%%%%%%%%%%%%%%%%%%%%%%%%%%%%%%%%%%%%%%%%%%%%%%

%%%%%%%%%%%%%%%%%%%%%%%%%%%%%%%%%%%%%%%%%%%%%%%%%%%%%%%%%%%%%%
\end{document}